\documentclass{elsart}
\journal{NIM}

\usepackage{amssymb}
\usepackage{amsmath,amsfonts,verbatim,graphicx,float}
\usepackage{subfigure}
\usepackage{cite}

\newfont{\tensy}{cmsy10}

\begin{document}
 
\begin{frontmatter}



\title{Tests of a Prototype Multiplexed Fiber-Optic Ultra-fast FADC Data 
Acquisition System for the MAGIC Telescope }

\author[munich]{H. Bartko\corauthref{cor1}},
\corauth[cor1]{Corresponding author.}
\ead{hbartko@mppmu.mpg.de}
\author[munich]{F. Goebel},
\author[munich]{R. Mirzoyan},
\author[munich]{W. Pimpl},
\author[munich]{M. Teshima}

\address[munich]{Max Planck Institute for Physics, F\"ohringer Ring 6, 80805 
Munich, Germany}

\begin{abstract}

Ground-based Atmospheric Air Cherenkov Telescopes (ACTs) are successfully used 
to observe very high energy (VHE) gamma rays from celestial objects. The light 
of the night sky (LONS) is a strong background for these telescopes. The gamma 
ray pulses being very short, an ultra-fast read-out of an ACT can minimize the 
influence of the LONS. This allows one to lower the so-called tail cuts of the shower image and the analysis energy threshold. It 
could also help to suppress other unwanted backgrounds.

Fast 'flash' analog-to-digital converters (FADCs) with GSamples/s are 
available commercially; they are, however, very expensive and power consuming. 
Here we present a novel technique of Fiber-Optic Multiplexing which uses a 
single 2 GSamples/s FADC to digitize 16 read-out channels consecutively. The 
analog signals are delayed by using optical fibers. The multiplexed (MUX) FADC 
read-out reduces the cost by about 85\% compared to using one ultra-fast FADC 
per read-out channel.

Two prototype multiplexers, each digitizing data from 16 channels, were built 
and tested. The ultra-fast read-out system will be described and the test 
results will be reported. The new system will be implemented for the read-out of the 17~m diameter MAGIC telescope camera. 

\end{abstract}


\end{frontmatter}

\section{Introduction}

MAGIC is the world-wide largest Imaging Air Cherenkov Telescope (IACT). It aims 
at studying gamma ray emission from the high energy phenomena and the violent 
physics processes in the universe, at the lowest energy threshold among existing 
IACTs. An overview about the gamma ray astronomy with IACTs is given in 
\cite{gamma-astro-status}. MAGIC is a unique detector in that 
it will cover the presently unexplored energy range between gamma ray satellite 
missions and other ground-based Cherenkov telescopes \cite{MAGIC-Proposal}. 

The camera of the MAGIC Telescope consists of 576 Photomultiplier tubes (PMTs), 
which deliver about 2 ns FWHM fast pulses to the experimental control house.
The currently used read-out system \cite{Magic-DAQ} is relatively slow (300 
MSamples/s). To record the pulse shape in detail, an artificial pulse stretching 
to about 6.5 ns FWHM is used. This causes more light of the night sky to be 
integrated, which acts as additional noise. Thus the analysis energy threshold 
of the telescope is limited, and the selection efficiency of the gamma signal 
from different backgrounds is reduced.

For the fast Cherenkov pulses (2~ns FWHM), a FADC with 2~GSamples/s can provide at 
least four sampling points. This permits a reasonable reconstruction of the 
pulse shape and could yield an improved gamma/hadron separation based on 
timing. Such an ultra-fast read-out can strongly improve the performance of 
MAGIC. The improved sensitivity and the lower analysis energy threshold will 
considerably extend the observation range of MAGIC, and allow one to search for very weak sources at high redshifts.

A few FADC products with $\geq 2$~GSamples/s and a bandwidth $\geq 500$~MHz are available commercially; they are, however, 
very expensive and power-consuming. To reduce the cost of an ultra-fast read-out 
system, a 2~GSamples/s read-out system has been developed at the Max-Planck-Institut f\"ur Physik in Munich. It uses the novel technique of Fiber-Optic 
Multiplexing, an approach possible because the signal duration (few ns) and 
the trigger frequency (typically $\sim$1~kHz) result in a very low duty cycle 
for the digitizer.
The new technique uses a single FADC of 700 MHz bandwidth and of 2~GSamples/s to digitize 16 read-out 
channels consecutively. The analog signals are delayed by using optical fibers. 
A trigger signal is generated using a fraction of the light, which is branched 
off by fiber-optic light splitters before the delay fibers. With the Fiber-Optic 
Multiplexing a cost reduction of about 85\% is achieved compared to using one 
FADC per read-out channel.

The suggested 2 GSamples/s multiplexed (MUX) FADC system will have a 10 bit 
amplitude resolution. For large signals the arrival time of the Cherenkov pulse 
can be determined with a resolution better than 200 ps. The system is relatively 
simple and reliable. All optical components and the FADCs are commercially 
available, while the multiplexer electronics has been developed at the MPI in 
Munich. Two prototype multiplexers, for 32 channels in total, were built and 
tested in-situ as read-out of the MAGIC telescope in La Palma in August 2004. 

In section \ref{sec:FADC_MAGIC} the MAGIC experiment is briefly described in the 
context of the data acquisition (DAQ) system using ultra-fast FADCs. The 
specifications of the ultra-fast read-out are described in section 
\ref{sec:MUX_description}, followed by the measured performance for the MUX-FADC 
prototype in laboratory tests (section \ref{sec:lab_tests}) and as read-out of 
the MAGIC telescope (section \ref{sec:LaPalma_test}).  Finally, section 
\ref{sec:discussion} is dedicated to discussions and conclusions.

\section{Principle and Signal Processing of the MAGIC Telescope} 
\label{sec:FADC_MAGIC}

\begin{figure}[h!]
\begin{center}
\includegraphics[totalheight=7cm]{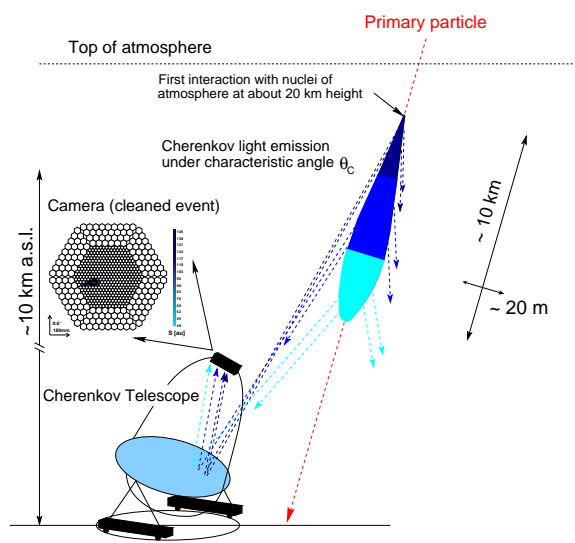}
\end{center}
\caption[IACT principle.]{\small \it IACT principle: A cosmic high energy gamma 
ray penetrates in the earth's atmosphere and initiates a shower cascade of 
electrons and positrons, which radiate Cherenkov light. This light is collected 
and focussed onto the camera, providing an image of the air shower. Picture 
taken from \cite{Sebastian}.} \label{fig:IACT_principle}
\end{figure}

Since the details of the MAGIC telescope are described elsewhere \cite{MAGIC-commissioning}, only items relevant to the FADC system are presented in this 
section. Figure \ref{fig:IACT_principle} shows the working principle of an air Cherenkov telescope. A 
high energy gamma ray entering the earth's atmosphere initiates a shower cascade 
of electrons and positrons. These radiate Cherenkov light, which is collected by 
the mirror and focussed onto the PMT camera of the MAGIC telescope. The main 
background originates from much more frequent showers induced by isotropic hadronic cosmic rays.

\begin{figure}[h!]
\begin{center}
\includegraphics[totalheight=8cm]{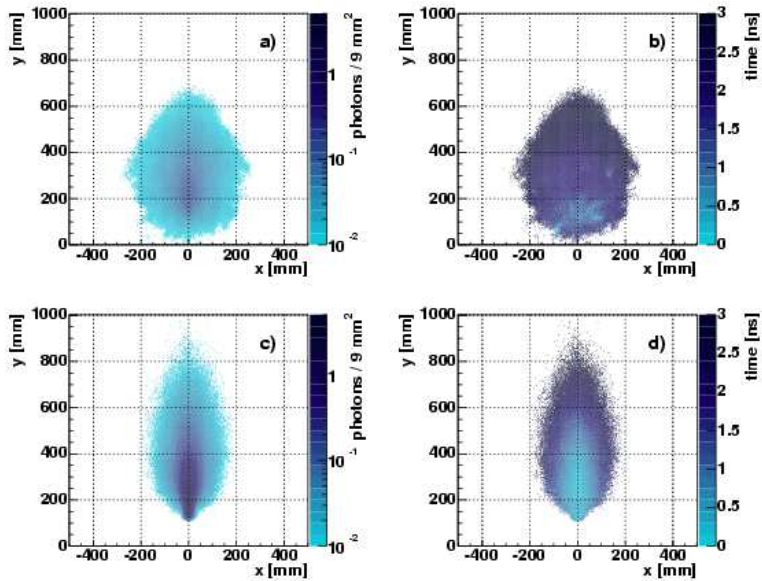}
\end{center}
\caption[Differences in timing structure for gammas and hadrons.]{ \small \it 
Mean amplitude (a, c) and time (b, d) profiles for gamma (c, d) and hadron (a, 
b) induced air showers images on the MAGIC camera plane from MC simulations. The impact parameter is fixed to 120 
m and the initial energy of the gamma is set to 100 GeV, while the proton energy 
is set to 200 GeV. The profiles are obtained by averaging over many simulated 
showers \cite{MC_shower_templates}. The timing structure of the image can 
provide viable information about the head and the tail of the shower as well as 
help to discriminate between gamma and hadron induced showers.} 
\label{fig:timing_gamma_had_differences}
\end{figure}

Monte Carlo (MC) based simulations predict different time structures for gamma 
and hadron induced shower images as well as for images of single muons. The 
timing information is therefore expected to improve the separation of gamma 
events from the background events \cite{muon_rejection}. Figure \ref{fig:timing_gamma_had_differences} 
shows the mean amplitude (a, c) and time (b, d) profiles for gamma (c, d) and 
hadron (a, b) induced air showers images on the camera plane of the MAGIC telescope. The impact parameter is fixed to 120 m and 
the initial gamma energy is set to 100 GeV, while the proton energy is set to 
200 GeV. The profiles are obtained by averaging over many simulated showers 
\cite{MC_shower_templates}. Although the total shower duration of gamma and hadron induced air showers is comparable, the photon arrival time varies smoothly over the gamma shower image while hadron shower images are structured in time. The time structure of the image can also provide 
essential information about head and tail of the shower.

\begin{figure}[h!]
\begin{center}
\includegraphics[width=\textwidth]{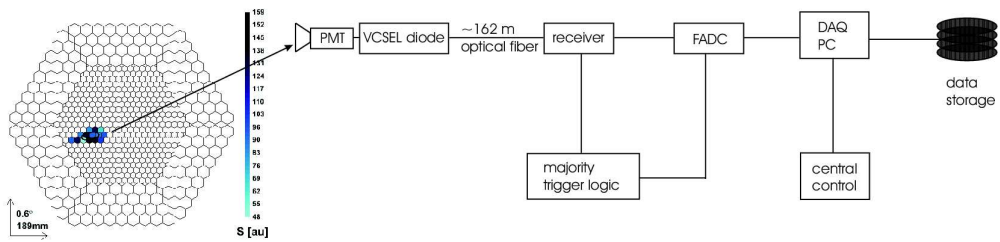}
\end{center}
\caption[Current MAGIC read-out scheme.]{ \small \it Current MAGIC read-out 
scheme: the analog PMT signals are transferred via an analog optical link to the 
counting house where after the trigger decision the signals are digitized by 
using a 300 MHz FADCs system and written to the hard disk of a DAQ PC.} 
\label{fig:MAGIC_read-out_scheme}
\end{figure}

The MAGIC read-out chain, including the PMT camera, the analog-optical link, the 
majority trigger logic and FADCs, is schematically shown in figure 
\ref{fig:MAGIC_read-out_scheme}. The response of the PMTs to sub-ns input light
shows a pulse of FWHM of 1.0 - 1.2 ns and rise and fall times of 600 and 700 ps 
correspondingly \cite{Magic-PMT}. By modulating vertical cavity surface emitting 
laser (VCSEL) diodes in amplitude the ultra-fast analogue signals from the PMTs 
are transferred via 162m long, multimode graded index 50/125 $\mu$m diameter 
optical fibers to the counting house \cite{MAGIC-analog-link-2}. After 
transforming the light back to an electrical signal, the original PMT pulse has 
a FWHM of about 2.2 ns and rise and fall times of about 1ns.

In order to sample this pulse shape with the present 300~MSamples/s FADC system, 
the pulse is stretched to a FWHM of $>6$ ns (the original pulse is folded with a 
stretching function of about 6ns). This implies a longer integration of LONS and thus 
the performance of the telescope on the analysis level is degraded. 

Because the current MAGIC FADCs have a resolution of 8 bit only, the signals are 
split into two branches with a factor of 10 difference in gain. One branch is 
delayed by 55 ns and then both branches are multiplexed and consecutively read 
out by one FADC. The FADC system can be read out with a maximum sustained rate 
of 1 kHz. A 512 kbytes FIFO memory allows short-time trigger rates of up to 50 
kHz.

\section{The Ultra-fast Fiber-Optic MUX-FADC Data Acquisition System} 
\label{sec:MUX_description}

The MAGIC collaboration is going to improve the performance of its telescope by 
installing a fast ($\geq2$~GSamples/s) FADC system, which fully exploits the 
intrinsic time structures of the Cherenkov light pulses. The requirements for 
such a system are the following:

\begin{itemize}
\item{10~bit resolution at a 2~GSamples/s sampling rate}
\item{$\geq500$ MHz bandwidth of the electronics chain including the FADC}
\item{up to 1~kHz sustained event trigger rate}
\item{dead time $\leq5\%$.}
\end{itemize}

\subsection{General MUX-Principle}

It is interesting to note that in experiments where FADCs are used to read out a 
multichannel detector in the common event trigger mode, only a tiny fraction of 
the FADC memory depth is occupied by the signal while the rest is effectively 
``empty'' \cite{MUX-priciple,MUX_ICRC}. One can try to correct this 
``inefficiency of use'' by ``packing'' the signals of many channels sequentially 
in time into a single FADC channel, i.e. by multiplexing. 

\begin{figure}[h!]
\begin{center}
\includegraphics[totalheight=7cm]{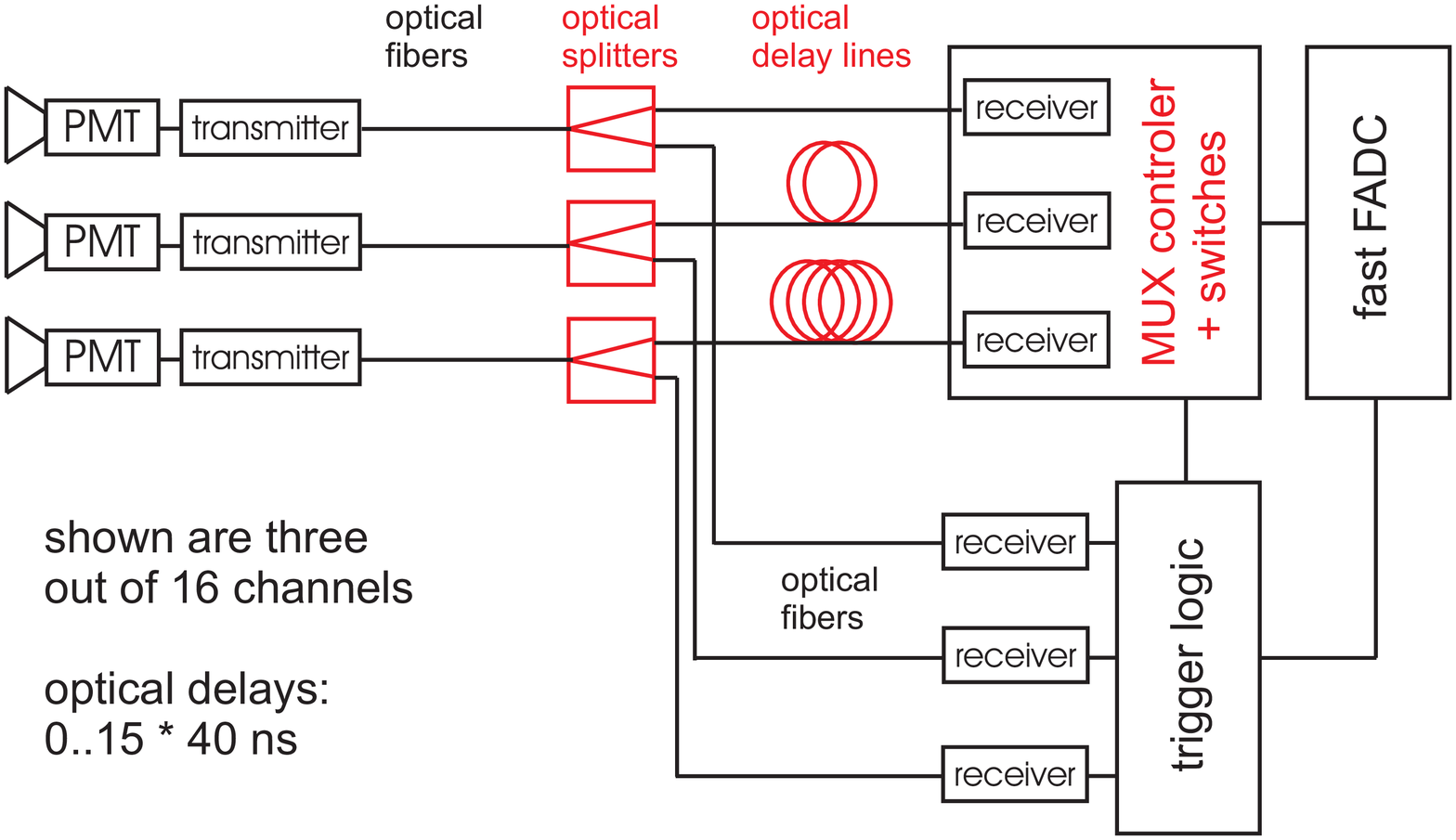}
\end{center}
\caption[The multiplexed FADC concept.]{ \small \it Schematic diagram of the 
multiplexed fiber-optic ultra-fast FADC read-out. Part of the analog signal that 
arrives via the fiber-optic link from the PMT camera is branched off and fed into 
a majority trigger logic. The other part of the signals is consecutively delayed 
in optical fibers. Channels are connected consecutively one by one to the 
ultra-fast FADC, using fast switches. Thereby the noise from the other channels 
is efficiently blocked.} \label{fig:mux_concept}
\end{figure}

Following this simple idea, a multiplexing system with fiber-optic delays has 
been developed for the MAGIC telescope. The block diagram is shown in figure 
\ref{fig:mux_concept}. The ultrafast fiber-optic multiplexer consists of three 
main components:

\begin{itemize}
\item{fiber-optic delays and splitters}
\item{multiplexer electronics: fast switches and controllers}
\item{ultra-fast FADCs.}
\end{itemize}

After the analog optical link between the MAGIC PMT camera and the counting 
house the optical signals are split into two parts. One part of the split signal 
is used as an input to the trigger logic. The other part is used for FADC 
measurements after passing through a fiber-optic delay line of a channel-specific length. 

The multiplexer electronics operates in the following way: The common trigger 
from the majority logic unit opens the switch of the first channel and allows 
the analog signal to pass through and be digitized by the FADC. All the other 
switches are closed during this time. When the digitization window for the first 
channel is over the corresponding switch is closed. The closed switch strongly 
attenuates the signal transmission by more than 60 dB for the fast MAGIC 
signals. Then the switch number two is opened such that the accordingly delayed 
analog signal from the second channel is digitized and so forth, one channel at 
a time until the last one is measured. In this way one ``packs'' signals from 
different channels in a time sequence which can be digitized by a single FADC 
channel. 

Because of the finite rise and fall times of the gate signals for the switches 
and because of some pick-up noise from the switch one has to allow for some 
switching time between the digitization of two consecutive channels. The gating 
time for each channel was set to 40 ns, of which the first and last 5 ns are 
affected by the switching process.

For the use in MAGIC a  $16 \rightarrow 1$ multiplexing ratio was chosen. 16 
channels are read out by a single ultra-fast FADC channel. The chosen 
multiplexing ratio is a compromise between

\begin{itemize}
\item{Dead time: the digitization of one event takes 16*40~ns=640~ns. During 
this time no other event can be recorded. Compared to the maximum sustained 
trigger rate of up to 1 kHz this dead time is negligible.}
\item{Noise due to cross-talk through the closed switches: The attenuated noise 
of the other channels could influence the active signal channel.}
\item{Cost of the FADCs}
\item{Mechanical constraints, e.g. board size, length of wires and fibers.}
\end{itemize}

\subsection{Optical Delays and Splitters}

Optical fibers were chosen for the analog signal transmission between the PMT 
camera and the counting house because they are lightweight, compact, immune to 
electro-magnetic pick-up noise and provide no pulse dispersion and attenuation 
\cite{MAGIC-analog-link-2}. The signal attenuation at 1 km fiber length is about 
2.3 dB for the chosen 850 nm wavelength of the VCSELs. The analog signal 
transmission offers a dynamic range larger than 60 dB.

Using fiber-optic delays ultra-fast analog signals can be delayed by several 
hundreds of ns. Thus a large number of successively delayed signals can be 
multiplexed and read out by a single channel FADC. Part of the analog signal has 
to be split off before the delay lines for
trigger. Therefore fiber-optic splitters of type $1 \rightarrow 2$ are used.

\begin{figure}[h!]
\begin{center}
\includegraphics[totalheight=7cm]{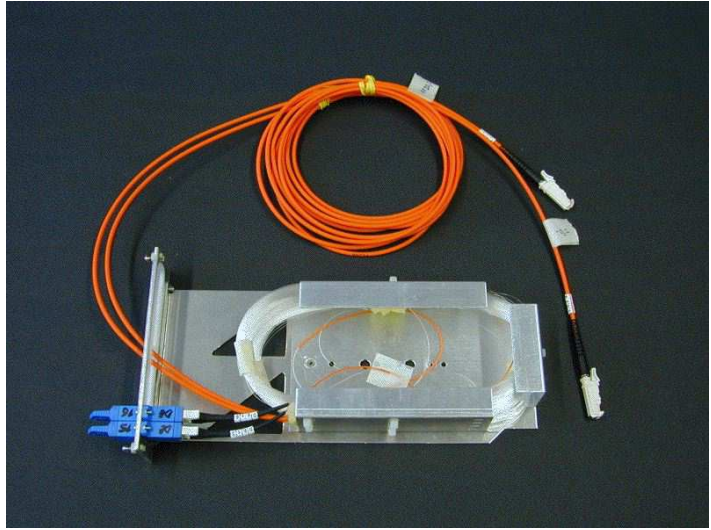} 
\end{center}
\caption[Two channel fiber-optic delay module.]{ \small \it Two channel fiber-optic delay module of 142 and 150 m length, corresponding to a delay of 710 and 
750 ns, respectively. Mechanical dimensions: 235 mm * 130 mm (3U) * 35 mm 
(7HP).} \label{fig:delay_oben}
\end{figure}

\begin{figure}[h!]
\begin{center}
\includegraphics[totalheight=7cm]{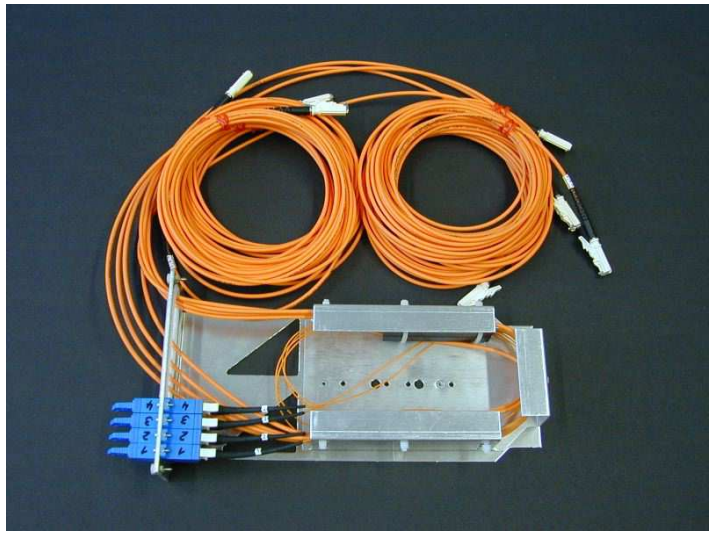}  
\end{center}
\caption[Four channel fiber-optic splitter module.]{ \small \it Four channel 
fiber-optic splitter module, GRIN technology and 50:50 splitting ratio. The 
outer dimensions are: 235 mm * 130 mm (3U) * 35 mm (7HP).} 
\label{fig:GRIN_splitter_module}
\end{figure}

Figure \ref{fig:delay_oben} shows a module containing two optical delay lines of 
142 m and 150 m length, corresponding to a delay of 710 ns and 750 ns. Figure 
\ref{fig:GRIN_splitter_module} shows a module of four graded index (GRIN)-type 
fiber-optic splitters with 50:50 splitting ratio (for a technical description 
see \cite{GRIN_splitters}). The modules have standardized outer dimensions and 
can be assembled in 3U hight 19'' crates. The splitters and optical delay lines 
are commercially available from the company Sachsenkabel \cite{Sachsenkabel}.

\subsection{MUX Electronics}

The multiplexer electronics consist of four stages. The first stage is a fiber 
optic receiver, where the signals from the optical delay lines are converted 
back to electrical pulses using PIN diodes. In a second stage, part of the 
electrical signal is branched off and transferred to a monitor output. The third 
stage consists of ultra-fast switches which are activated one at a time. In the 
last stage all 16 channels are summed to one output. The multiplexed signals are 
then transferred via 50 $\Omega$ coaxial cables to the FADC channels. Table 
\ref{table:MUX_board_specifications} summarizes the specifications of the 
multiplexer electronics.

\begin{table}[h]{\normalsize\center
\caption{\small \it Specifications of the electronics for analog signal 
multiplexing.}\label{table:MUX_board_specifications}}
\begin{tabular}{ll}
 \hline
 Mechanical size & 370 mm (9 U) * 220 mm * 30 mm (6 HP)
\\ Number of channels & 16
\\ Analog input & via 50/125 $\mu$m graded index fiber, E2000 connector 
\\ Gain & 25, including the VCSEL transmitter
\\ Dynamic range & max output amplitude: 1 V
\\ Power supplies & +12 V, $\pm5$ V
\\ Power dissipation & $\sim$20 W 
\\ Trigger input & LVDS\\ 
\hline
\end{tabular}
\end{table}

\begin{figure}[h!]
\begin{center}
\includegraphics[totalheight=7cm]{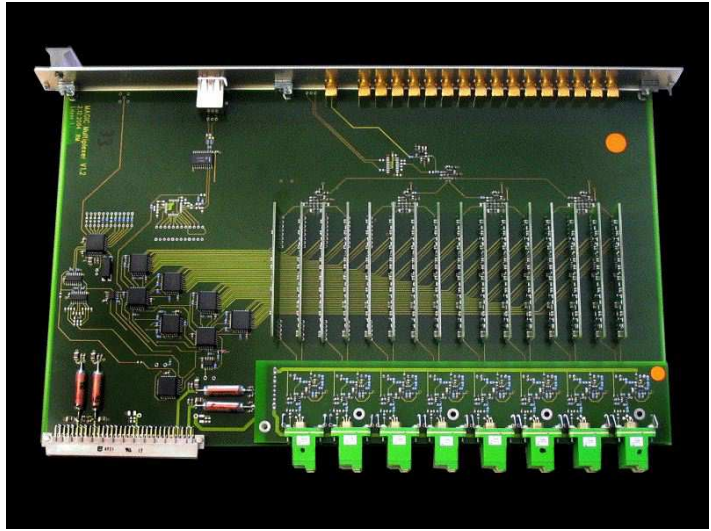}
\end{center}
\caption[The printed circuit board for analogue signal multiplexing.]{ \small 
\it Photo of the printed circuit board for analogue signal multiplexing: It 
consists of a trigger input, 16 opto-electric converters, 16 monitor signal 
outputs, the Digital Switch Control circuit (DSC), 16 daughter switch boards and 
two summing stages. The overall size is 370 mm (9 U) * 220 mm * 30 mm (6 HP).} 
\label{fig:neues_mux_board}
\end{figure}

One multiplexer module consists of one 6 layer {\it motherboard} and 16 double 
layer {\it switchboards}, which are plugged into the {\it motherboard} via 
multiple pin connectors. Figure \ref{fig:neues_mux_board} shows a photo of the 
printed circuit MUX {\it motherboard} with 16 mounted daughter {\it 
switchboards}.

The {\it motherboard} includes the following components:

\begin{itemize}
\item{16 opto-electrical converters}
\item{16 monitor outputs}
\item{the Digital Switch Control circuit (DSC)}
\item{the trigger input to activate the DSC}
\item{16 ultra-fast switches on 16 {\it switchboards}}
\item{the 16 channel summing stage.}
\end{itemize}

One opto-electrical converter consists of a receptacle, a PIN photo diode, 
packed in the E2000-connector. The photodiode is biased by 12 V to reduce its 
intrinsic capacity for speed and noise optimization. The current signal of the 
PIN photo diode is converted into an equivalent voltage signal by a 
transimpedance-amplifier. Its amplifier-IC has a gain-bandwidth product of about 
1.5 GHz and a very high slew rate of about 4000 V/$\mu$s. The trans-impedance is 
1000 $\Omega$. A monitor output consists of an ultra-wide band (UWB)-driver-amplifier, which transmits the signal from the transimpedance-stage to a 50 
$\Omega$-SMA-connector.

\begin{figure}[h!]
\begin{center}
\includegraphics [totalheight=9cm]{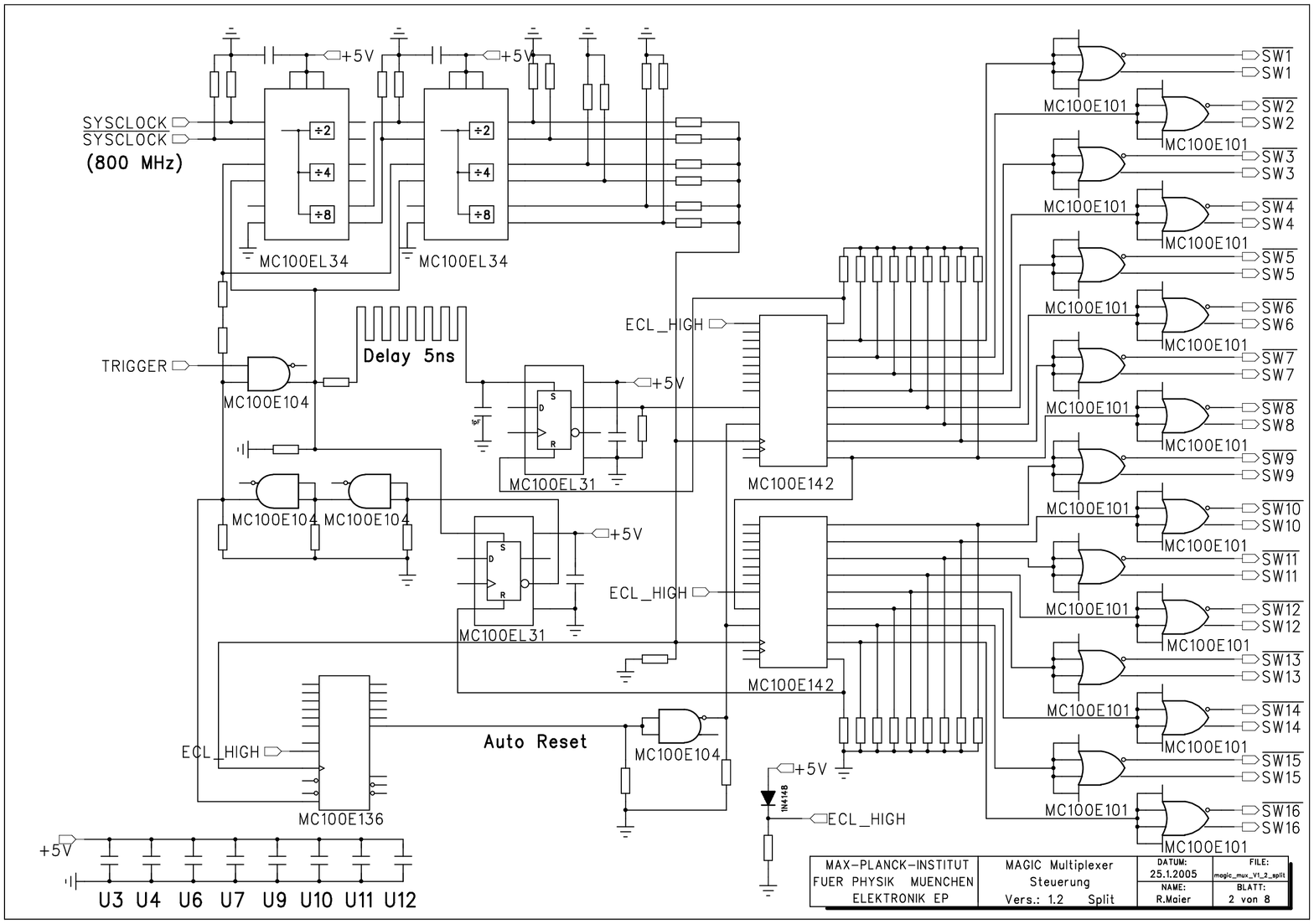}
\end{center}
\caption[Circuit diagram of the Digital Switch Control circuit (DSC).]{ \small 
\it Circuit diagram of the Digital Switch Control circuit (DSC): The trigger 
initiates a sequence of 16 PECL high levels of 40 ns duration applied 
consecutively to the switch boards.} \label{fig:mux_electronics_scheme}
\end{figure}

Figure \ref{fig:mux_electronics_scheme} shows the circuit diagram of the Digital 
Switch Control circuit, DSC. It consists of the following parts:

\begin{itemize}
\item{One clock generator-IC. It is programmable with a resolution of 12 bit 
from 50 MHz to 800 MHz and works in PECL mode. It is crystal stabilized and set 
to 800 MHz.}
\item{A digital delay line (DDL) that can be set from 2 ns to 10 ns with 11 bit 
accuracy. It can be used to adjust the trigger times between different MUX {\it 
motherboards}.}
\item{A digital lock-in-circuit (DLC) synchronizes the MUX-sequence to the 
trigger signal. The lock-in jitter is 1.25 ns ($=1/[800~\mathrm{MHz}]$).}
\item{16 differential PECL-drivers that transmit the MUX-sequence signals to the 
corresponding {\it switchboards}.}
\end{itemize}

\begin{figure}[h!]
\begin{center}
\includegraphics[totalheight=7cm]{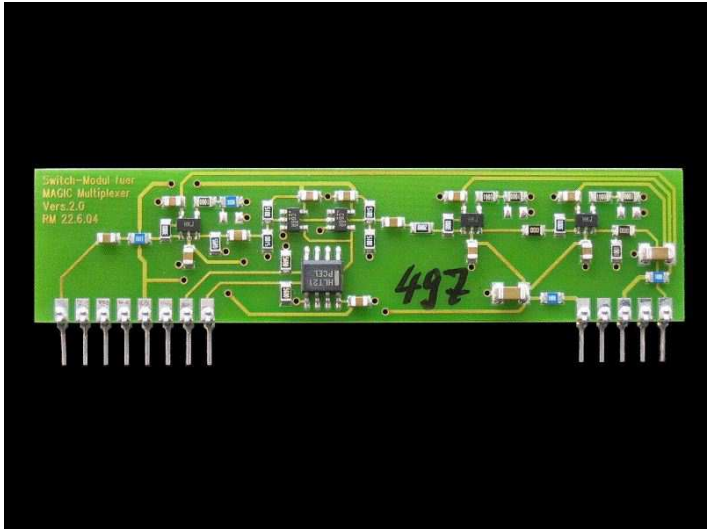} 
\end{center}
\caption[The printed circuit board for fast switching.]{ \small \it The printed 
circuit board for fast switching. The switch board contains two ns-MOSFET-switches operated in series. Its mechanical dimensions are 80 mm * 20 mm *5 mm.} 
\label{fig:switch_board}
\end{figure}

\begin{figure}[h!]
\begin{center}
\includegraphics[totalheight=8cm]{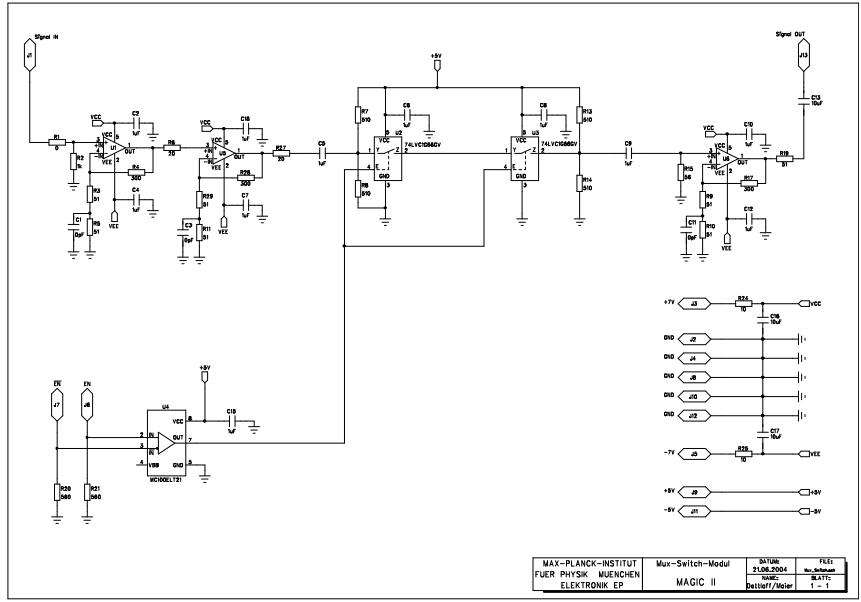}
\end{center}
\caption[Schematics of the fast switches.]{ \small \it Schematics of the fast 
switches: A high PECL level from the Digital Switch Control circuit opens both 
of the ns-MOSFET switches operated in series.} \label{fig:switch_scheme}
\end{figure}

Each switchboard includes two ultra-wideband (UWB)-amplifier circuits, followed 
by two ns-switching MOSFETs operated in series and one UWB-driver-amplifier-circuit. MOSFET switches were chosen due to their fast switching properties and 
a very fast stabilization of the signal baseline after the switching. The small 
cross-talk through the closed switch is further reduced by the serial operation 
of two switches. An on-board PECL to CMOS converter distributes the digital 
switch-control-circuit (DSC)-signal to the MOSFET-switches in parallel. Figure 
\ref{fig:switch_board} shows a photo of the switch board, while its circuit 
diagram is shown in figure \ref{fig:switch_scheme}.

\begin{figure}[h!]
\begin{center}
\includegraphics[angle=-90,totalheight=9cm]{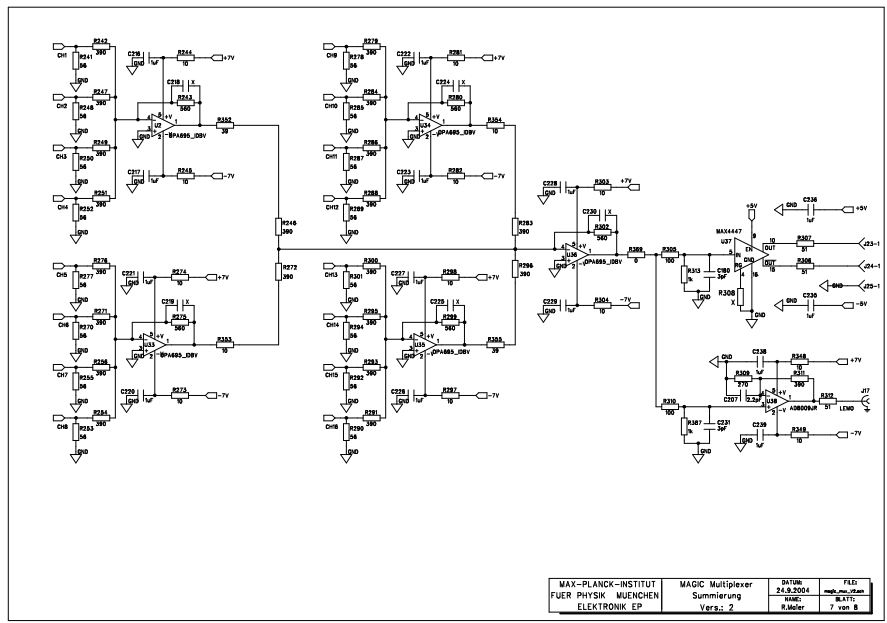}
\end{center}
\caption[Schematics of the summing stage.]{ \small \it Schematics of the 
two-step summing stage of the multiplexed signals. The two-step setup keeps the 
channel-wires short and allows to use the UWB-amplifiers in the faster inverting 
mode while keeping the signal polarity non-inverting.} \label{fig:MAGIC-Mux_schematic7}
\end{figure}

In a passive summation, the switch parasitic capacitances would add up and 
can significantly widen the signal pulse. To avoid this, a two-step 
active summation was chosen: In the first step, the outputs of four channels are 
summed together. In the second step, the four resulting outputs are summed into 
one. For the summing UWB-amplifiers are used. The two-step setup keeps the 
channel wires short, and permits to use the amplifiers in the faster inverting 
mode while keeping the signal polarity non-inverting. Finally, an UWB-driver 
sends the multiplexed signals over a 50 $\Omega$-SMA-coaxial connection to the 
FADC. The circuit diagram of the summation stage is shown in figure 
\ref{fig:MAGIC-Mux_schematic7}.

\subsection{FADC Read-Out}

The FADCs are commercial products manufactured by the company Acqiris (DC 282) 
\cite{Acqiris}. They feature a 10 bit amplitude resolution, a bandwidth of 700 
MHz, a sampling speed of 2 GSamples/s and an input voltage range of 1 V. Each 
FADC board contains 4 channels. The read-out data are stored in a RAM on the 
FADC board of 256 kSamples (512 kbytes) size per channel. Up to 4 FADC boards 
can be arranged in one compact PCI crate and are read out by a crate controller 
PC running under Linux. The FADCs are designed for a 66 MHz 64 bit data transfer 
via the compact PCI bus.

The FADC features a trigger time interpolator TDC that can be used to correct 
for a potential trigger jitter of 500 ps due to the asynchronous FADC clock with 
respect to the trigger decision. Table \ref{table:FADC_specifications} 
summarizes the specifications of the ultra-fast FADCs.

\begin{table}[h]{\normalsize\center
\caption{\small \it Specifications of the ultra-fast 
FADC.}\label{table:FADC_specifications}}
\begin{tabular}{ll}
  
\hline
 Mechanical size & 267 mm (6 U) * 220 mm * 30 mm (6 HP)
\\ Number of channels & 4
\\ Analog input & 1 V full scale, adjustable offset
\\ Sampling frequency & 2 GSamples/s
\\ Sampling resolution & 10 bits
\\ RAM size & 256 kSamples (512 kbytes) per channel
\\ Bandwidth & 700 MHz
\\ Noise level & $<1.2$ LSB guaranteed
\\ Power dissipation & 60 W (4 channels)
\\ Trigger input & unipolar, adjustable threshold\\ 
\hline
\end{tabular}
\end{table}

\section{Performance of the System Components} \label{sec:lab_tests}

The performance of the MUX-FADC system components was studied in extensive 
laboratory tests. The quality and performance of the FADCs and of different 
commercially available optical splitters and delays were evaluated. Several 
iterations of the multiplexer electronics design were made.

\subsection{Performance of the Optical Delays and Splitters}

The fiber-optic delay lines have channel-specific delay times of 0...15 times 
40~ns plus 500~ns common base delay. Deviations from the specified delay times 
and potential changes in the delay due to temperature variations are important. 
It has to be ensured that all signals arrive in time at the multiplexer 
electronics when a given switch is open. 

The manufacturer guarantees for the delay length a maximum deviation of 
$\pm$2~ns from its nominal value. This was confirmed in laboratory measurements. Although the signal attenuation in fibers is small, nevertheless, there are small differences in the dispersion of the signals in different channels due to the different delay lengths.


Different technologies of fiber-optic splitters are available on the market. 
Three splitting technologies were tested: fused splitters, bifurcation splitters 
and so called GRIN-splitters. In the fused technology two optical fibers are 
drilled and then thermally fused together. In bifurcation splitters the end 
faces of the two output fibers are mechanically attached to the end face of the 
input fiber. In GRIN-type splitters the splitting is done by a semi-transparent 
mirror in conjunction with two graded index lenses \cite{GRIN_splitters}.

The MAGIC optical link uses multimode VCSELs and multimode optical fibers. 
Mechanical stress or deformations of the input fiber into the splitter, 
especially due to telescope movements, can vary the propagation of light modes in the fiber. 
The expected movements of the fibers were simulated in the laboratory by bending 
the fibers using different bending radii. The fused and bifurcation fiber-optic 
splitters show changes in the splitting ratio of more than $\pm$10 \%. Only the 
so called GRIN type splitters are immune against mode changes, with measured changes of 
the splitting ratio of less than 1\%.

The splitting ratio is guaranteed to be 50:50 within $\pm3$ \% by the 
manufacturer. This was again confirmed in test measurements. All tested 
splitters were found to be insensitive with respect to time and temperature 
changes.

\subsection{Performance of the MUX Electronics}

The MUX electronics was extensively tested in the laboratory. For the use as a 
read-out system for MAGIC, the following points are very important: 

\begin{itemize}
\item{short switching noise and flat signal base-line}
\item{high bandwidth (low pulse dispersion in amplitude and in time)} 
\item{strong signal attenuation for closed switches}
\item{good linearity and large dynamic range}
\item{stability.}
\end{itemize}

\begin{figure}[h!]
\begin{center}
\includegraphics[angle=-90, totalheight=7cm]{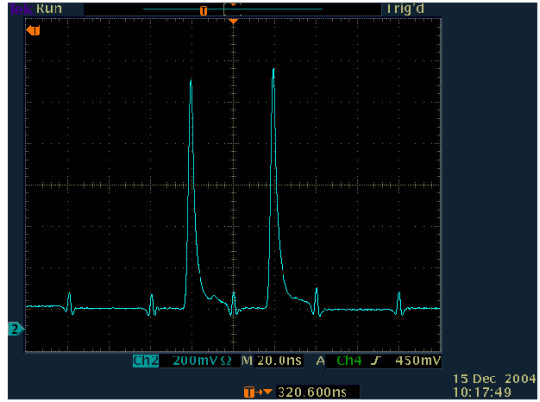}%
\end{center}
\caption[Oscilloscope snapshot of two consecutive multiplexed signals. Two 
additional signal baselines can be seen.]{ \small \it Oscilloscope photo of two 
consecutive multiplexed signals and the baseline of two empty signal gates.} 
\label{fig:cmos_2signale_neu}
\end{figure}

Figure \ref{fig:cmos_2signale_neu} shows a photo recorded with a fast 
oscilloscope of two consecutively multiplexed signals along with the switching 
noise between two channels. Although the switching noise is as large as 100 mV, 
it is very reproducible and confined to less than 10 ns of the 40 ns window per 
read-out channel. In the rest of the window the baseline is flat and stable.

The switching and summing stages only slightly widen the fast input pulses. A 
pulse of about 2.5 ns FWHM after the receiver photo diode is widened to 2.7 ns 
FWHM at the output of the multiplexer electronics. The pass-through of such fast 
signals through closed switches is less than 0.1\% (60 dB attenuation).

\begin{figure}[h!]
\begin{center}
\includegraphics[totalheight=6cm]{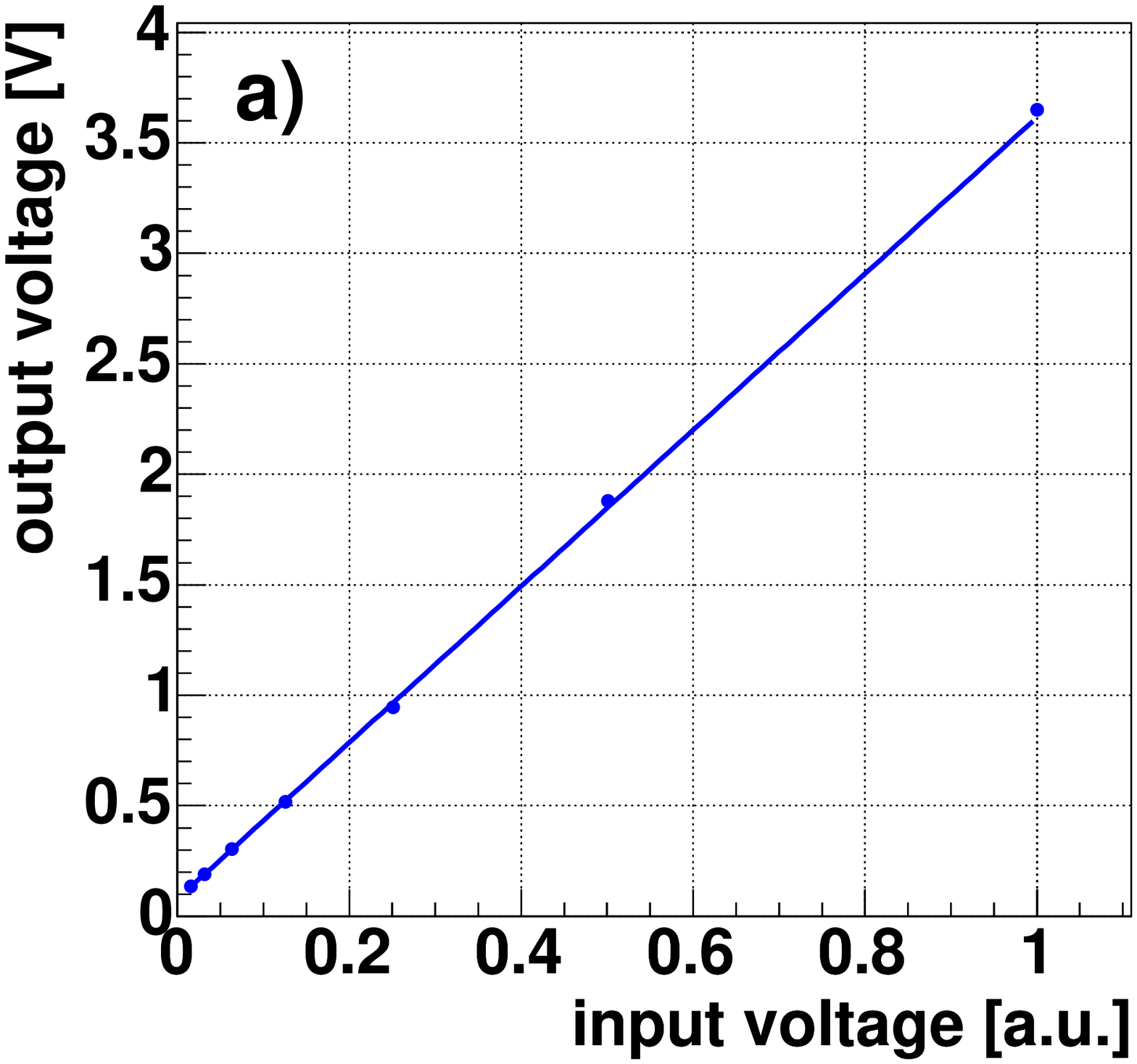}
\includegraphics[totalheight=6cm]{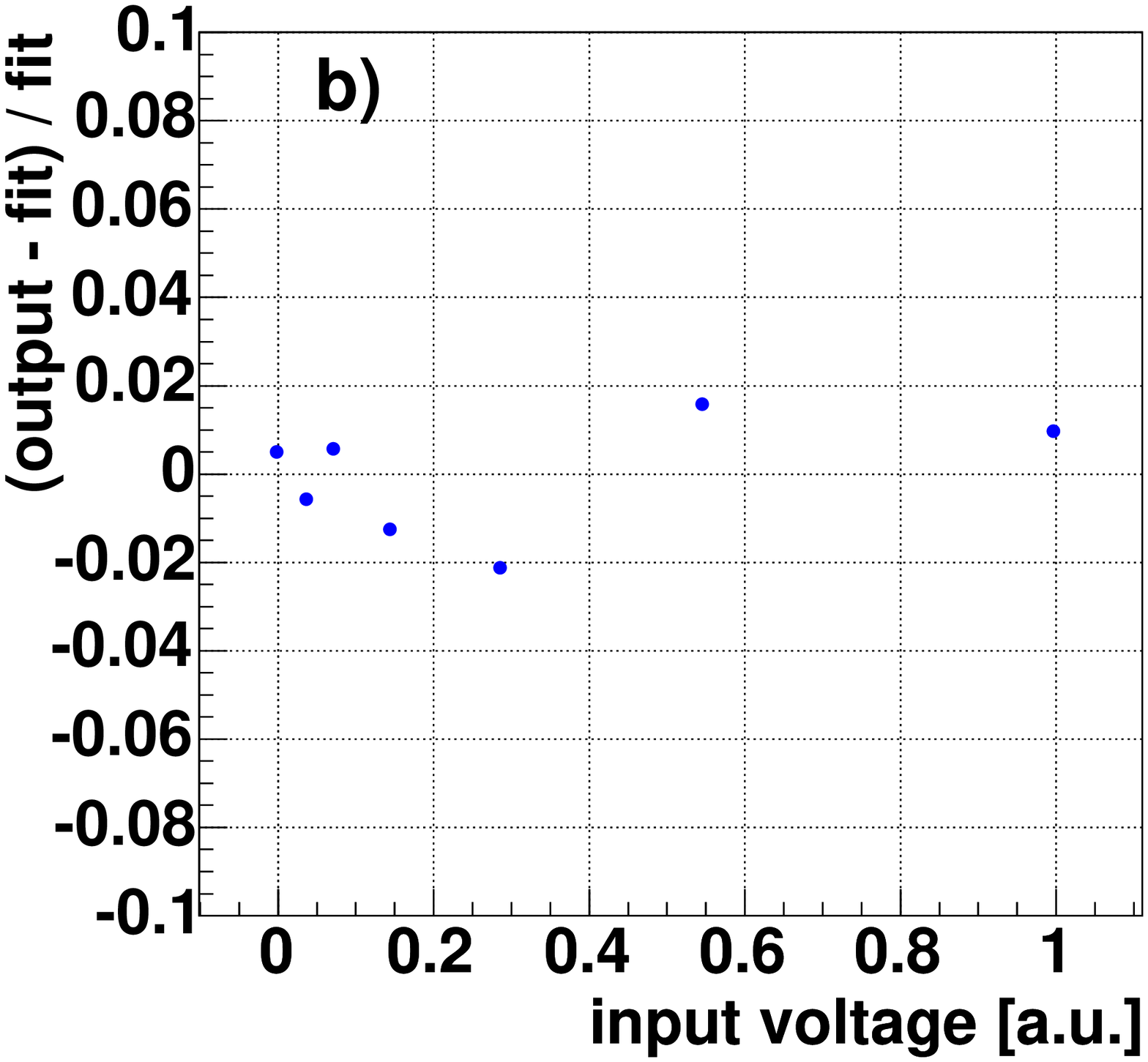}
\end{center}
\caption[Linearity of the switches.]{ \small \it a) Output signal amplitude of 
the MUX board as a function of the input signal after the PIN diode as measured 
at the monitor output. b) Residua of the linearity of the MUX board as a 
function of the input signal after the PIN receiver diode as measured at the 
monitor output. The deviations are less than 2\%. } 
\label{fig:linearity}
\end{figure}

Figure \ref{fig:linearity}a shows the combined linearity of the switches and of 
the summing stage. The output signal amplitude of the MUX-board is plotted as a 
function of the input signal amplitude after the PIN-Diode, as measured at the 
monitor output. The right panel of the figure shows the deviations from 
linearity of the MUX-electronics. For output signals up to 1 V the MUX-
electronics is linear with differential deviations less than 2\%. The total non-linearity of the read-out chain is dominated by the analog optical link. Its response deviates from a perfect linear behavior by less than 10\% in a total range of 56 dB \cite{Magic-analog-link-linearity}.

\subsection{FADC Performance}

The main performance parameters of the FADC are 

\begin{itemize}
\item{noise level / effective dynamic range}
\item{linearity and bandwidth}
\item{maximum trigger and acquisition rate}
\item{dead time.}
\end{itemize}

A noise level of less than 1.2 least significant bits (LSBs) is guaranteed by 
the manufacturer (700 MHz bandwidth, no input amplifier). There are small but 
constant differences in the input voltage full scales and thus in the gain for 
different FADC channels. These can be corrected for by the offline calibration 
software. The FADCs feature an internal calibration system keeping their 
integral and differential non-linearity below one LSB.

For the maximum trigger rate and the dead time optimization the interplay 
between the FADC boards and the crate controller PC is important. The compact 
PCI (cPCI) bus allows an effective data throughput of up to about 400 Mbytes/s (66 MHz, 64 bit) shared between all FADC channels in 
one crate. 

In each event 2560 bytes are stored per FADC channel (16 channels of 40 ns 
gate time, 2~GSamples/s and 2~bytes per sample for the 10-bit resolution FADC). 
Reading out 8~FADC channels with one crate controller board results in a data 
volume of about 20~kbytes/event, which has to be transferred via the cPCI bus. 

The ultra-fast FADC offers three modes of data acquisition:

\begin{itemize}
\item{single acquisition}
\item{segmented memory}
\item{asynchronous acquisitions using a FIFO memory.}
\end{itemize}


In the single acquisition mode the FADC writes the digitized data into the on-board RAM using it as one big ring buffer. Upon the arrival of a trigger the $N = 2560$ bytes corresponding to the event are copied to the PC with a DMA transfer via the cPCI bus. The read-out time $T_1$ per event and per 
FADC channel in the crate is given by the sum of the DMA overhead time, 
$\mathrm{Ovhd}_{\mathrm{DMA}} \leq 25 \mu$s and the data transfer time over the 
cPCI bus \cite{Acqiris}:
\begin{equation}
T_1 = \mathrm{Ovhd}_{\mathrm{DMA}} + N \cdot \mathrm{Xfr} 
\ .
\end{equation}
$\mathrm{Xfr} = 2.5 \mathrm{ns/byte}$ is the data throughput of the cPCI bus for the 64 bit, 66 MHz operation. 

For 8 FADC channels per crate this amounts to a total transfer time of about $250 \mu$s. Including a $\leq 25 \mu$s  global trigger rearm time this leads to about $275 \mu$s dead time per event. 



In the segmented mode the FADC RAM is divided into many segments. Each
segment is used as a circular buffer where the digitized data is stored.
After the arrival of a trigger the digitizer continues to write into the
next segment. The dead time between two events is $\leq 25\mu$s.
When all segments are filled the data is copied to the PC via
one DMA transfer.
The total readout time for M segments is:

\begin{equation}
T_{2} = \mathrm{Ovhd}_{\mathrm{DMA}} +  M \cdot (N+\mathrm{Extra}) \cdot \mathrm{Xfr} \ , 
\end{equation} 

where $\mathrm{Extra} \leq 200$ denotes the number of ``overhead'' data points 
per segment. 

In the current scheme an additional time to reorder the data inside the PC
has to be taken into account. For 8 FADC channels in one crate and 100
segments the total dead time amounts to $\leq 16 $ms, i.e. $\leq 160 \mu$s per event.



The most attractive operation mode uses the FADC RAM as a FIFO. In this
case the FADC writes the digitized data in one part of the RAM while
previously stored data is asynchronously transfered to the PC. The dead
time is thus reduced to the $\leq 25 \mu$s needed to rearm the trigger.
Additional dead time arises only if the average trigger rate exceeds the
maximum sustainable rate of 2 kHz.


\section{Prototype Test in the MAGIC Telescope on La 
Palma}\label{sec:LaPalma_test}

Two prototype MUX-FADC read-out modules for 32 channels were tested as a read-out of the MAGIC telescope during two weeks in August/September 2004.

The main goals for the tests were:

\begin{itemize}
\item{test of concept of the ultra-fast MUX-FADCs under realistic conditions}
\item{study the interplay of the MUX-FADC system with the MAGIC trigger and data 
acquisition system}
\item{implement the reconstruction and calibration for the ultra-fast digitized 
signals in the common MAGIC software framework MARS \cite{Magic-software}}
\item{provide input for detailed MC simulations for the ultra-fast 
digitization.}
\end{itemize}

\subsection{Setup of the Prototype Test}\label{sec:LaPalma_test_setup}

\begin{figure}[h!]
\begin{center}
\includegraphics[totalheight=7cm]{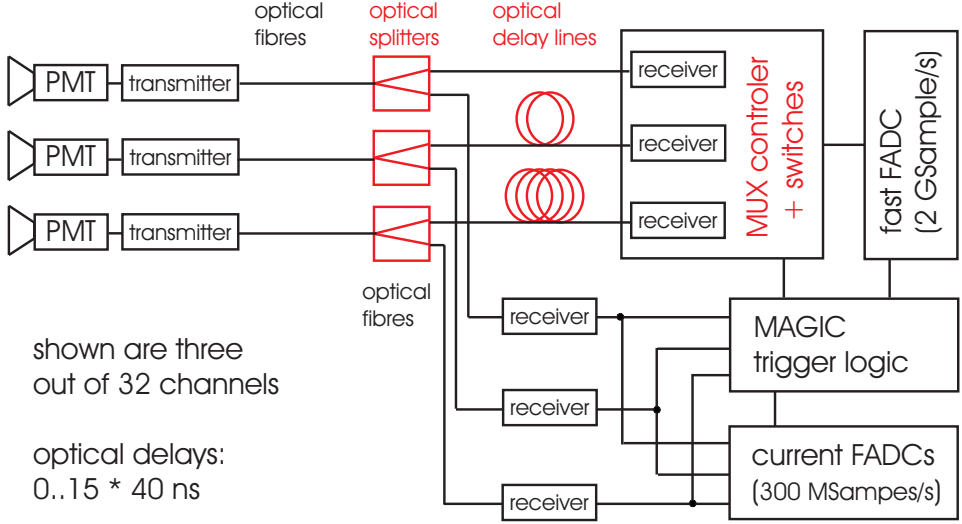}
\end{center}
\caption[Integration of the MUX-FADC read-out in the current MAGIC FADC read-
out.]{ \small \it Block diagram of the integration of the MUX-FADC prototype 
read-out in the current MAGIC FADC read-out. Shower images are simultaneously 
recorded with the ultra-fast MUX-FADC system and with the current MAGIC FADCs. 
The MAGIC trigger logic provides a trigger for the MUX electronics as well as 
for the ultra-fast FADCs.} \label{fig:mux_LaPalma_test_1}
\end{figure}

Two MUX boards of 16 channels each were integrated into the MAGIC read-out 
system allowing the  simultaneous data taking with the current 300 MSamples/s 
read-out and the MUX-FADC prototype read-out. Figure 
\ref{fig:mux_LaPalma_test_1} shows in a block diagram how the MUX-FADC prototype 
read-out system was integrated into the current MAGIC FADC read-out. The analog 
optical signals arriving from the MAGIC PMT camera were split into two equal 
parts using fiber-optic splitters. One part of the optical signal was connected 
to the current MAGIC receiver boards which provided output signals to the MAGIC 
majority trigger logic \cite{Magic-trigger} and to the current 300 MSamples/s 
FADCs. The other part of the optical signal was delayed by a channel specific 
delay of 0...15 times 40 ns plus common base delay and directed to the optical 
receivers on the MUX boards. The common MAGIC trigger was used to trigger the MUX boards as well as the fast FADCs.

Figure \ref{fig:event95} shows the group of 32 selected channels of the MAGIC 
PMT camera \cite{MAGIC-commissioning} to be read out by the ultra-fast 
digitizing system. The channels were chosen to be close packed in order to 
contain (at least partially) images of showers. In the test 16 bifurcation and 
16 GRIN type splitters were used.


In order to acquire only events where the shower image is located in the 32 MUX-FADC channels, only these channels were enabled in the MAGIC trigger system. The 
trigger fires if the signal in at least four close packed pixels exceeds the 
preset threshold.

In the prototype tests on La Palma an older version of the ultra-fast FADC was 
used, the Acqiris DC 240. It features a sampling speed of 2 GSamples/s with an 8 
bit resolution. It was connected via a PCI bridge to a host PC running under 
Windows.

For every trigger 1300 FADC samples (16 times 80 samples plus 20 extra samples) 
were recorded with both of the used multiplexed FADC channels. An FADC memory of 
120 segments was used. In the host PC the data were written into a binary file. 
This setup was chosen for simplicity and was not optimized for the smallest dead 
time in a continuous data taking mode. Nevertheless, the dead time between two 
of the 120 consecutively recorded events in the segmented mode was negligible.

\subsection{The Data}\label{sec:data_taken}

The tests of the MUX-FADC system were carried out around the full moon period. In 
total about 230000 triggers were taken with the MUX-FADC read-out system 
(including pedestals and calibration LED light pulses). Table 
\ref{table:data_taken} summarizes the amount of data taken with and without the 
presence of moon light. 

\begin{table}[h]{\normalsize\center
\caption{\small \it Overview of the data taken during the MUX-FADC prototype 
test in the MAGIC telescope at La Palma.}\label{table:data_taken}}
\begin{tabular}{l|r|r}
   trigger type &  current FADC read-out & MUX-FADC read-out 
\\ \hline
& & 
\\ pedestals, no moon    & 500    & 26400
\\ pedestals, moon       & 5000   & 13210
\\ calibration, no moon  & 47000  & 96000
\\ calibration, moon     & 91000  & 70420
\\ cosmics, no moon      & 500    & 8040
\\ cosmics, moon         & 0      & 16800  \\
\hline total                 & 144000 & 230870 \\ 
\end{tabular}
\end{table}

\subsection{Data analysis}\label{sec:data_analysis}

Each data file contains 120 events of two ultra-fast FADC channels with 1300 
recorded FADC samples per event. The recorded raw data are converted into the 
usual ROOT-based MAGIC raw data format \cite{Magic-software}, which provides the 
flexibility to adjust the number of recorded samples for each pixel.

\subsubsection{Signal Reconstruction}

\begin{figure}[h!]
\begin{center}
\includegraphics[totalheight=7cm]{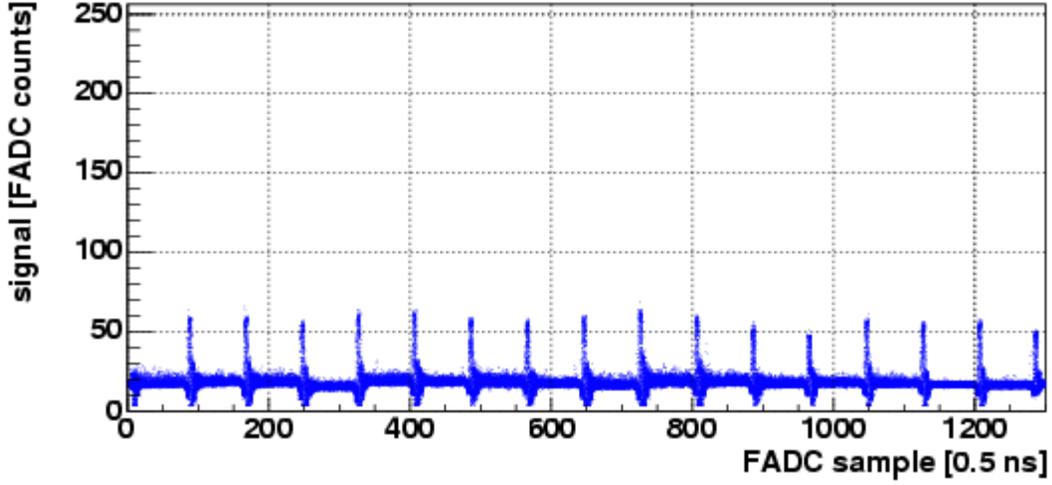}
\end{center}
\caption[Pedestals: Raw data.]{ \small \it Pedestals: Raw data (1300 samples for 
16 consecutive channels) for 120 randomly triggered events overlayed.} 
\label{fig:raw_FADC_ped_open}
\end{figure}

For each event the signals of 16 PMTs of the MAGIC camera are sequentially 
digitized by one FADC channel. As an example, figure \ref{fig:raw_FADC_ped_open} 
shows the raw data for 120 superimposed randomly triggered pedestal events. 
Between every two consecutive channels the switch noise is visible. 

\begin{figure}[h!]
\begin{center}
\includegraphics[totalheight=7cm]{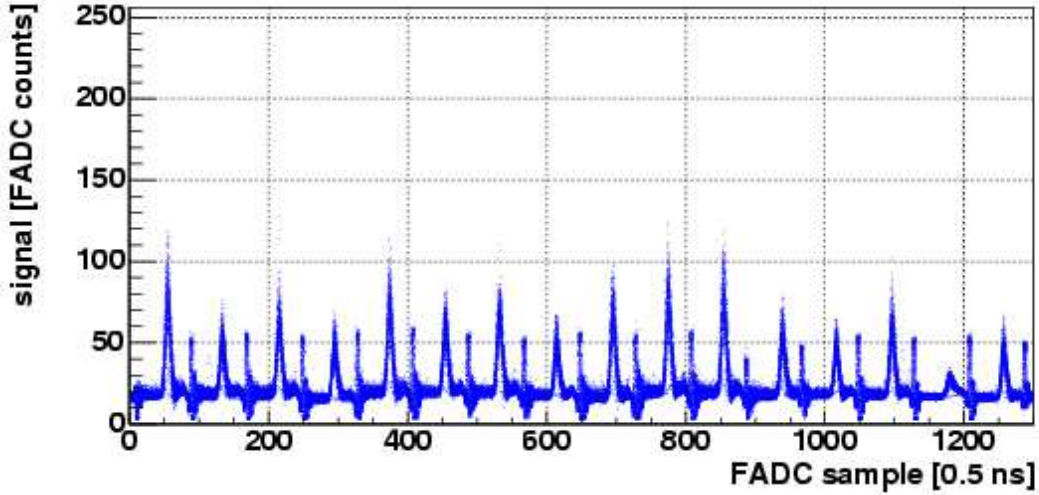}
\end{center}
\caption[Calibration: Raw data.]{ \small \it Calibration events: Raw data (1300 
samples for 16 consecutive channels) for 120 LED light pulses overlayed.} 
\label{fig:raw_FADC_13UV}
\end{figure}

For calibration purposes the MAGIC PMT camera can be uniformly illuminated by a 
fast LED light pulser located in the center of the telescope dish \cite{MAGIC-calibration}. Figure \ref{fig:raw_FADC_13UV} shows the raw data of 16 
consecutively read-out channels for 120 superimposed calibration events. The 
calibration signal pulses are clearly visible on the signal baseline. The gain 
difference from channel to channel is mainly due to a spread in the gain of the 
VCSEL and receiver diodes of the analog optical link. The additional spread due 
to small differences in the fiber-optic splitters and a signal attenuation in 
the delay lines is small.

For each channel the pedestal level and pedestal RMS are calculated from either 
a pedestal run with random triggers or directly from the data. For the pedestal 
calculation a fixed number of FADC samples at a fixed position in the 
digitization window is used.

For the signal reconstruction a fixed number of FADC samples is integrated. The 
integration interval was chosen to be 4 FADC samples (corresponding to 4*3.33 ns 
= 13.33 ns) for the current MAGIC FADCs. For the MUX-FADCs a window size of 10 
FADC samples is chosen, corresponding to a 5 ns integration window. The 
reconstructed signal $\overline{S}$ is then given by:
\begin{equation}
\overline{S}=\sum_{i=i_0}^{i=i_0+3(9)}{S_i} \ ,
\end{equation}
where $S_i$ is the i-th FADC sample after the trigger. The signal arrival time 
relative to the first FADC sample after the trigger, $t_{\text{arrival}}$, is 
reconstructed as the first moment of the FADC time samples used to calculate the 
reconstructed signal: 
\begin{equation} \label{time_extractor}
t_{\text{arrival}}=\frac{\sum_{i=i_0}^{i=i_0+3(9)}{S_i (t_i-
t_{i_0})}}{\sum_{i=i_0}^{i=i_0+3(9)}{S_i}} \ .
\end{equation}

\subsubsection{Calibration}

The calibration system of the MAGIC telescope consists of intensity controlled 
fast LED light pulsers of different colors and intensities that illuminate the 
MAGIC camera homogeneously \cite{MAGIC-calibration}. Using the laboratory 
measured excess noise factor of the MAGIC PMTs the conversion constant between 
reconstructed signals in FADC counts and photo electrons can be determined. The 
common MAGIC calibration algorithms and software were successfully applied to 
the ultra-fast digitization.

\begin{figure}[h!]
\begin{center}
\includegraphics[totalheight=7cm]{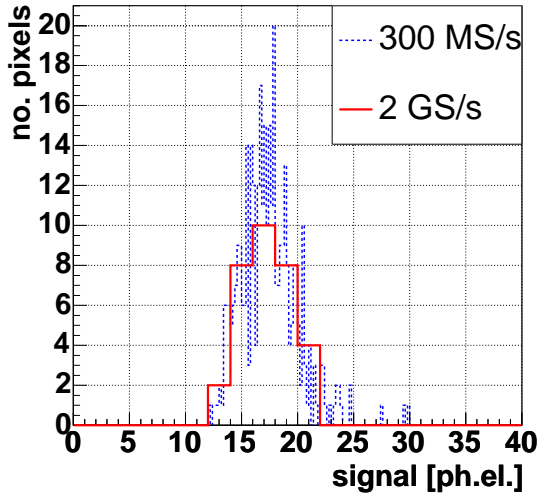}	
\end{center}
\caption[Mean reconstructed number of photo electrons for the LED pulser.]
{\small \it Distributions of the mean reconstructed number of photo electrons in 
the PMTs of the MAGIC camera from the LED pulser for the current 300 MSamples/s 
FADCs and the MUX-FADCs. Both read-out systems yield the same average number of 
photo electrons.} \label{fig:cam_phel_oldFADC}
\end{figure}

Figure \ref{fig:cam_phel_oldFADC} shows the distribution of the mean number of 
photo electrons per pixel reconstructed with the current 300 MSamples/s FADC 
system and the MUX-FADC system. The MAGIC camera was illuminated with UV 
calibration pulses of about 2.5 ns FWHM. As expected, the mean reconstructed number of photo 
electrons is the same for the 32 split channels used in the MUX-FADC tests as 
for all the other MAGIC read-out channels. 

Small differences in the cable length of the MAGIC analog optical link, the 
fiber-optic delays and transition times in the PMTs introduce arrival time 
differences between of the pulses in different read-out channels of up to a few 
ns. These relative channel to channel time differences can also be calibrated 
using the LED pulser. One can determine the mean time difference between all 
pixels with respect to a reference pixel. In the calibration procedure of the 
cosmics events this timing difference is corrected for.

\begin{figure}[h!]
\begin{center}
\includegraphics[totalheight=7cm]{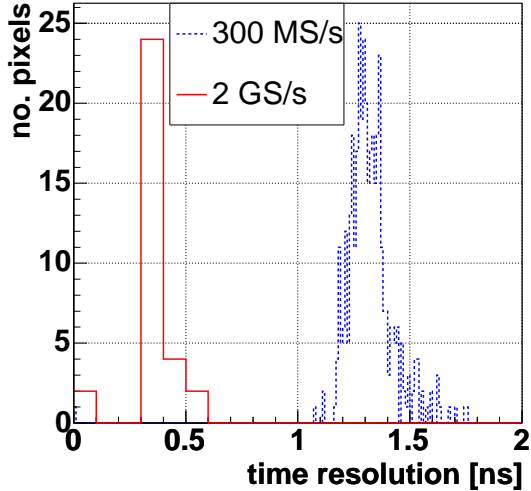}
\end{center}
\caption[Timing resolution determined for the LED pulser.]{\small \it 
Distributions of the timing resolution for the current 300 MSamples/s FADC read-out and the MUX-FADC read-out. The MUX-FADC system yields an improvement in the 
timing resolution by more than a factor of three.} \label{fig:time_resolution}
\end{figure}

In addition, the event to event variation of the timing difference between two 
read-out channels for the LED pulser provides a measure of the timing accuracy. 
Figure \ref{fig:time_resolution} shows the distributions of the determined 
timing resolution of the current 300 MSamples/s FADCs together with the timing 
resolution of the MUX-FADCs. The timing accuracy strongly depends on the signal 
to noise ratio and the width of the input light pulse. The MUX-FADCs yield a better timing resolution by more than a 
factor of three compared to the current FADC system using the simple and stable timing extraction algorithm of equation (\ref{time_extractor}).

\subsubsection{Cosmics Data}

Cosmics shower data were recorded to study in detail the interplay of the ultra-fast MUX-FADC system with the MAGIC trigger logic. It also provides valuable 
input for the MAGIC MC simulations of the ultra-fast digitization system, e.g. 
about the pulse shapes for cosmics events.

\begin{figure}[h!]
\begin{center}
\includegraphics[totalheight=6cm]{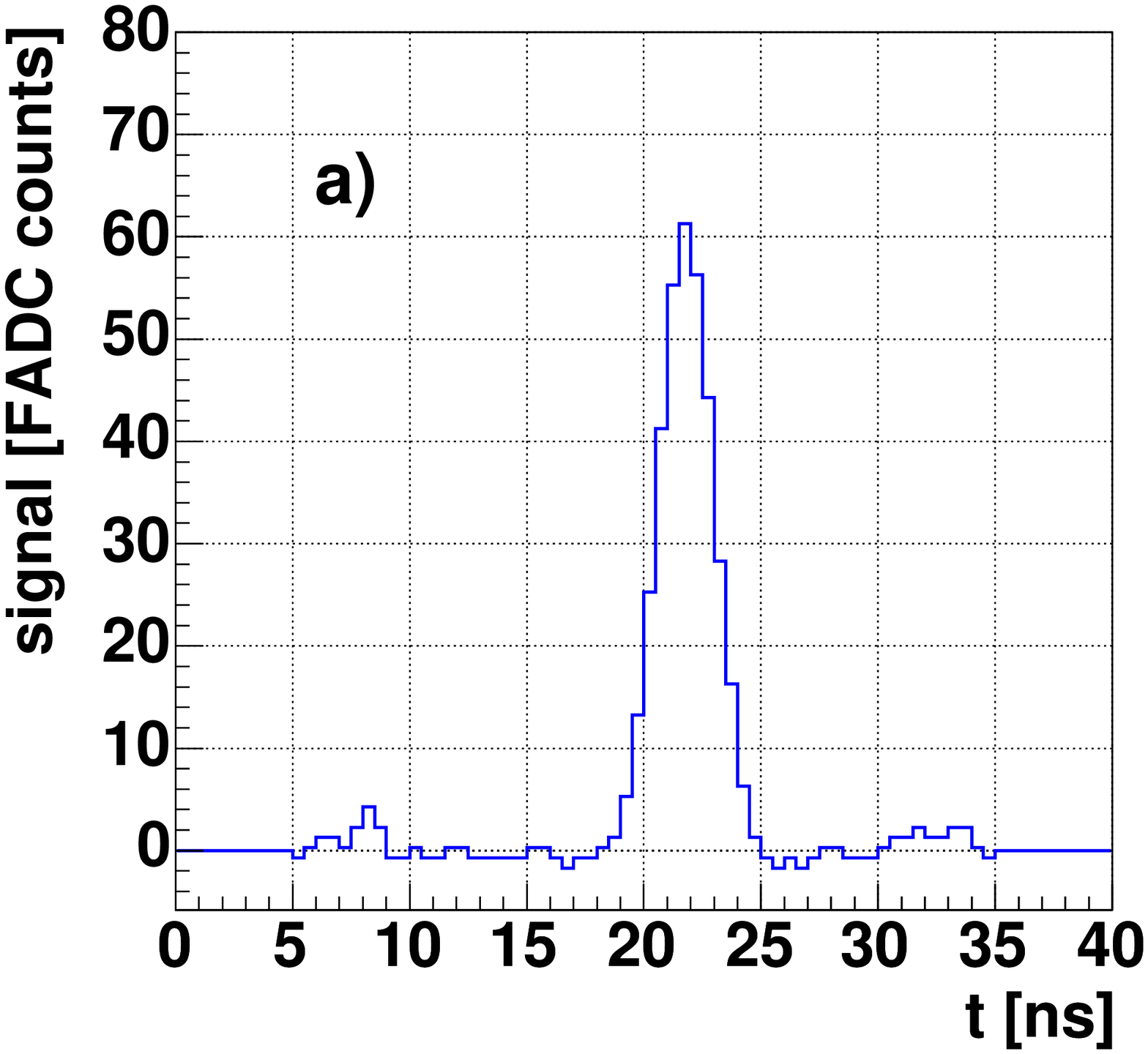}
\includegraphics[totalheight=6cm]{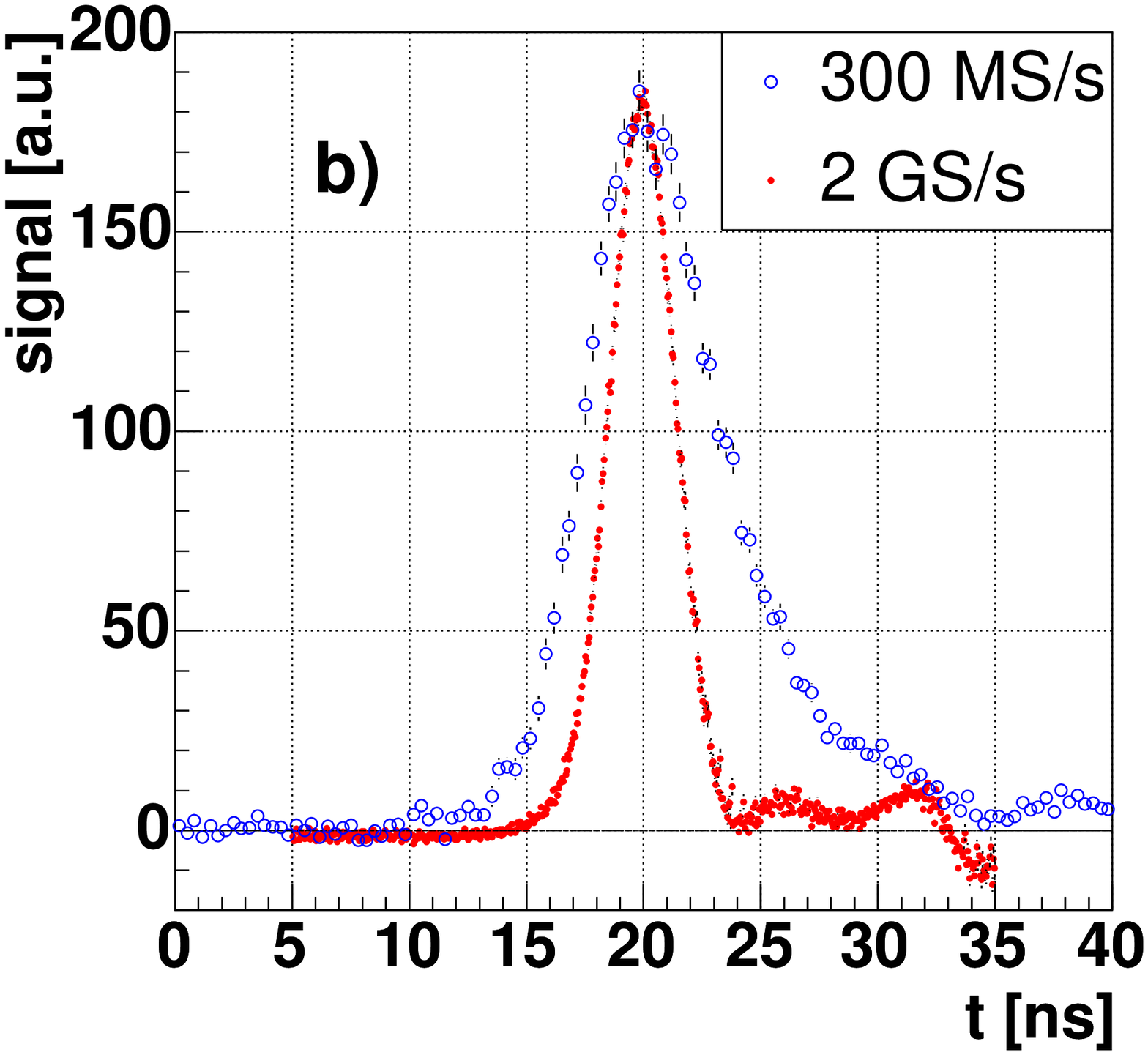}
\end{center}
\caption[Pulse shapes for cosmics events.]{\small \it a) Pulse shape in a single 
pixel for a typical cosmics event after pedestal subtraction. b) Comparison 
between the mean reconstructed pulse shapes recorded with the current MAGIC 
FADCs (open circles) and with the MUX-FADCs (full points).} 
\label{fig:shape_event95}
\end{figure}

In figure \ref{fig:shape_event95}a one can see  the pulse shape in a single 
pixel for a typical cosmics event. By overlaying the recorded FADC samples of 
many events after adjusting to the same arrival time, the average reconstructed 
pulse shapes can be calculated. Figure \ref{fig:shape_event95}b shows the 
comparison of the average reconstructed pulse shapes recorded with the current 
300 MSamples/s MAGIC FADCs, including the 6ns pulse stretching, and with the 
MUX-FADCs. The average reconstructed pulse shape for cosmics events has a FWHM 
of about 6.3 ns for the current FADC system and a FWHM of about 3.2 ns for the 
MUX-FADC system.

\begin{figure}[h!]
\begin{center}
\includegraphics[totalheight=7cm]{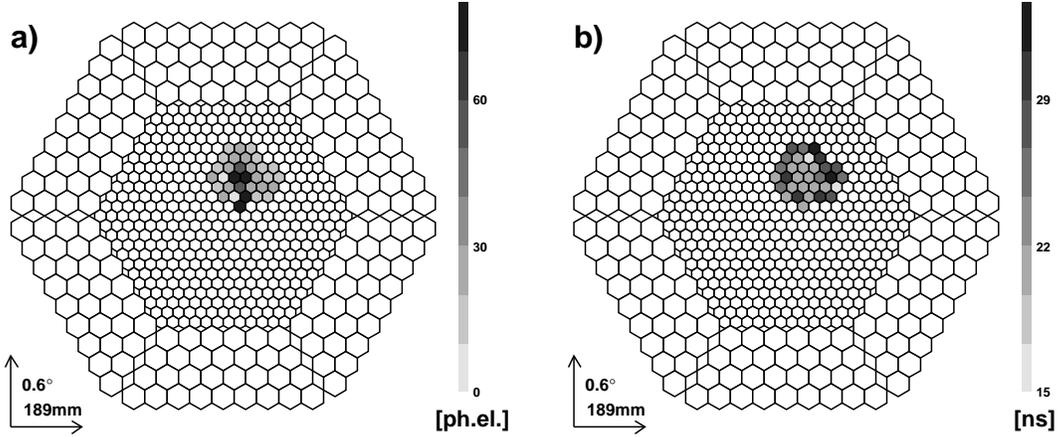}
\end{center}
\caption[Reconstructed signal and arrival times for a typical event.]
{\small \it a) Reconstructed signal in photo electrons in the MAGIC PMT camera 
display and b) calibrated arrival times in ns in the MAGIC camera display for a 
typical cosmics event.} \label{fig:event95}
\end{figure}

Figure \ref{fig:event95}a shows a MAGIC PMT camera display with the 
reconstructed signal after calibration in photo electrons for a typical cosmics 
event. For the same event figure \ref{fig:event95}b shows the reconstructed 
arrival time after correction for the cannel-to-channel time differences.

\subsubsection{Pedestals / Noise}

In the data recorded by an IACT, the electronics noise together with the LONS 
fluctuations is superimposed on the Cherenkov signal from air-showers. The noise 
from the LONS can be simulated as the superposition of the detector response to 
single photo electrons, arriving at a given rate but randomly distributed in 
time. This can be quantified using the noise autocorrelation function 
$\boldsymbol{B_{ij}}$, the correlation between the read-out samples $i$ and $j$:
\begin{equation}
\boldsymbol{B_{ij}} = \langle b_i b_j \rangle - \langle b_i \rangle \langle b_j 
\rangle  \ ,
\end{equation}
where $b_{i}$ and $b_{j}$ are the FADC samples $i$ and $j$ for a pedestal event.

\begin{figure}[h!]
\begin{center}
\includegraphics[totalheight=7cm]{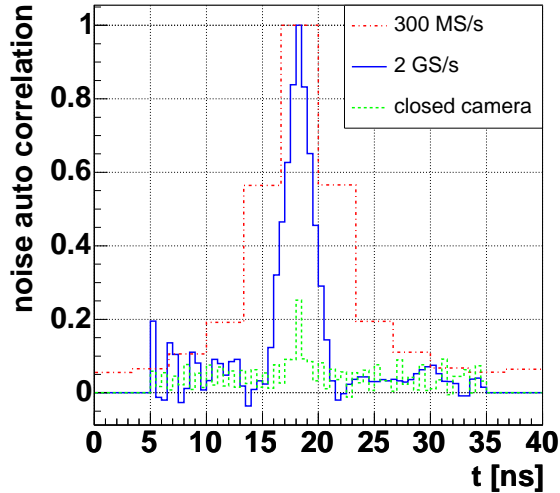} 
\end{center}
\caption[Noise autocorrelation.]{\small \it Noise autocorrelation function with 
respect to a fixed FADC sample for the current MAGIC readout chain with 6 ns 
pulse shaping, the MUX-FADC read-out with open camera and closed camera, 
normalized to the pedestal RMS of the opened camera.} \label{fig:autocorr_new}
\end{figure}

Figure \ref{fig:autocorr_new} shows the noise autocorrelation for the current 
MAGIC FADC system and the MUX-FADC system with open camera (i.e. exposed to LONS), normalized to the 
pedestal RMS. In the same plot, the noise autocorrelation for the MUX-FADC 
system with closed camera (no LONS), normalized to the pedestal RMS for an open camera, is 
shown. The noise autocorrelation of the current FADC system extends to several 
ns since the pulse is stretched by 6 ns. For the MUX-FADC system with no 
pulse shaping there is still a considerable noise autocorrelation 
for an open camera. The noise autocorrelation mostly disappears in case of a closed camera with electronics noise only. 

\begin{figure}[h!]
\begin{center}
\includegraphics[totalheight=7cm]{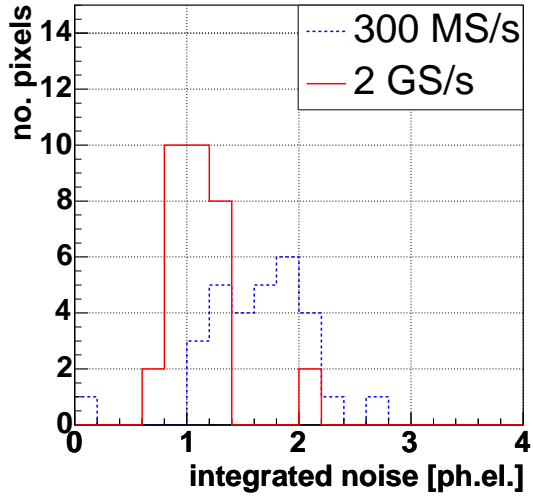}
\end{center}
\caption[Integrated noise after calibration into photo electrons.]
{\small \it Distributions of the integrated noise per pixel in the signal reconstruction 
window after calibration into photo electrons for the current 300 MSamples/s 
FADC read-out and using the MUX-FADC read-out.} \label{fig:cam_noise_oldMUX}
\end{figure}

Figure \ref{fig:cam_noise_oldMUX} shows the distributions of the integrated noise (integrated pedestal RMS
after calibration in photo electrons) per pixel for the current FADC system and for the 
MUX-FADC system. The shorter integration time used for the pulse reconstruction 
with the MUX-FADC system yields a reduction of the effective integrated noise by 
about 40\%. 

\begin{figure}[h!]
\begin{center}
\includegraphics[totalheight=7cm]{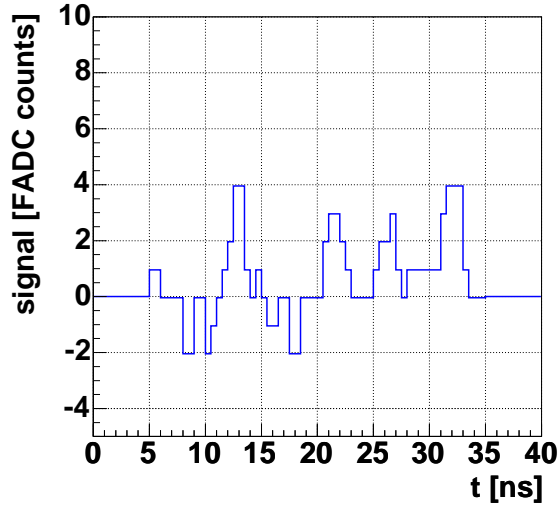}
\end{center}
\caption[Time structure in a pedestal event.]{\small \it Time structure in a 
typical pedestal event. The peaks on the baseline might be due to single photo 
electrons from the light of the night sky.} \label{fig:pedestal_vs_time}
\end{figure}

Using the new MUX-FADC system the noise contributions due to the LONS may even 
be resolved into individual pulses. Figure \ref{fig:pedestal_vs_time} shows a 
typical example for the signals in a pedestal event (random triggers). The 
pedestal does not vary in an uncorrelated way. Instead most of the pedestal 
variations are due to peaks on the signal baseline.

\begin{figure}[h!]
\begin{center}
\includegraphics[totalheight=6cm]{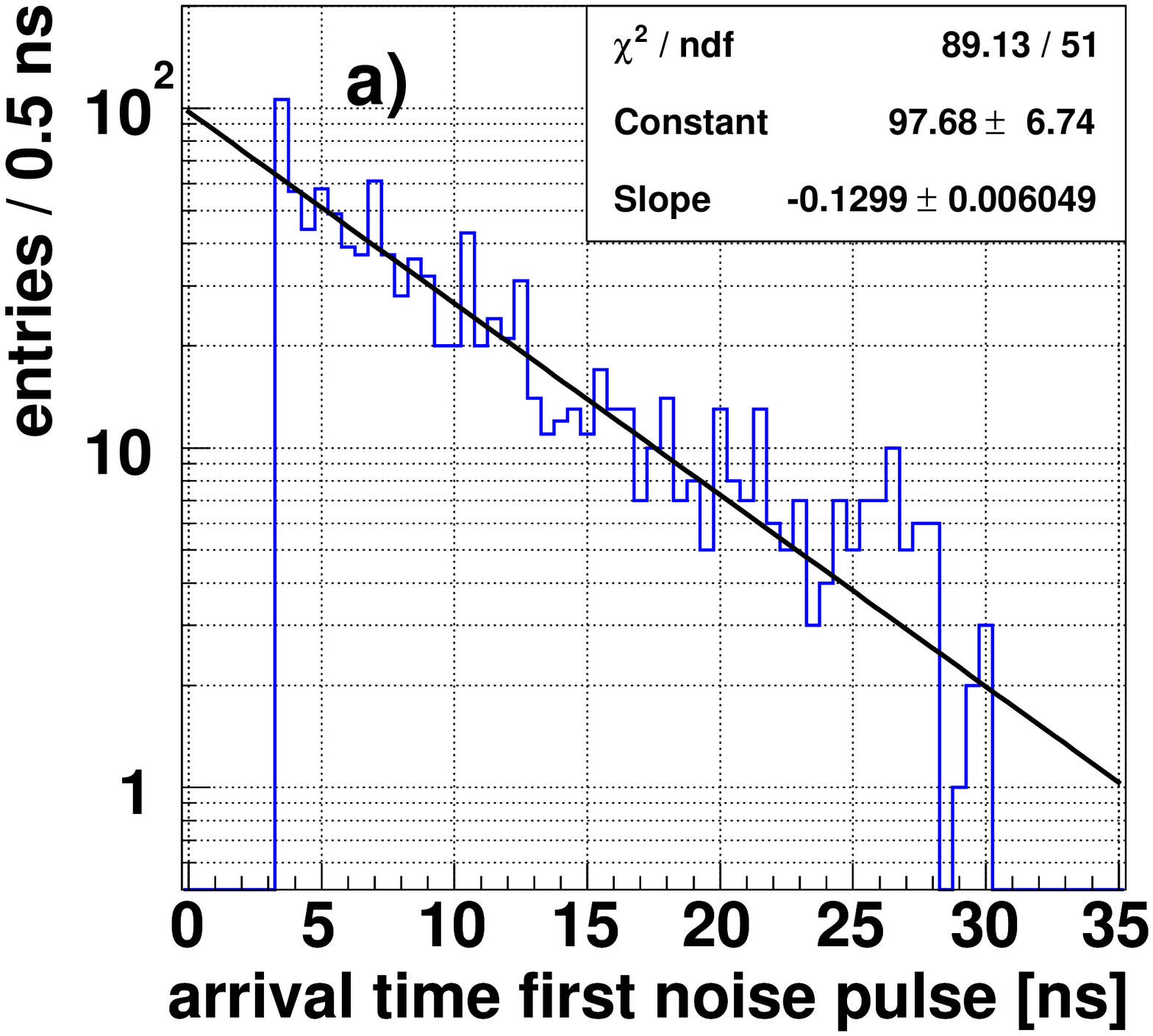}
\includegraphics[totalheight=6cm]{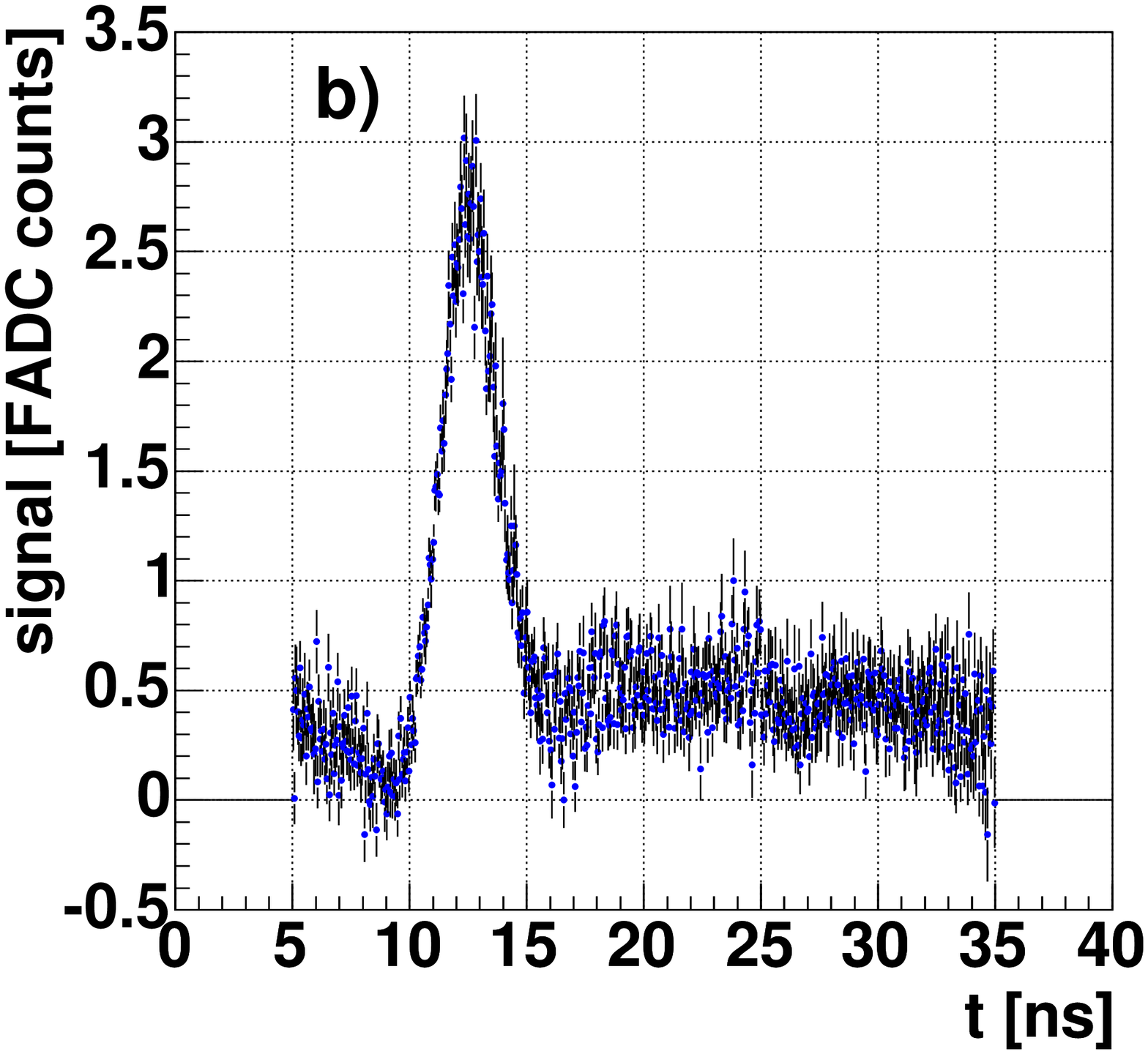}
\end{center}
\caption[Time distribution of the structures on the pedestal baseline.]
{\small \it a) Arrival time distribution of the first noise peak on the pedestal 
baseline. The peaks are arriving randomly in time with a rate of $(0.13 \pm 
0.01) \mathrm{ns}^{-1}$. b) Average reconstructed shape of the LONS noise 
peaks. The FWHM is about 2.6~ns.} \label{fig:noise_arrival_time_int_3}
\end{figure}

The rate of the peaks was studied to verify whether it is compatible with the 
rate of LONS photo electrons. A window of 6 slices is slid over the FADC samples 
of randomly triggered pedestal events. The first window position after the 
switch noise where the sum of the FADC samples exceeds the pedestal level by at 
least 3 FADC counts was chosen. Figure \ref{fig:noise_arrival_time_int_3}a shows 
the arrival time distribution of the first noise peak. The distribution can be 
fit by an exponential function with a rate $r$ of 
\begin{equation}
r=(0.13 \pm 0.01) \mathrm{ns}^{-1} \ .
\end{equation}
This corresponds to an integrated LONS charge of about 1.3 photo electrons per 10 ns 
integration window, which is in good agreement with the expected LONS rate.

Figure \ref{fig:noise_arrival_time_int_3}b shows the pulse shape of the selected 
noise peak averaged over many events. It has a FWHM of about 2.6~ns. This corresponds to the response of the instrument to a $\delta$-function input pulse (single LONS photo electrons have no internal time structure). The mean charge of the noise peak corresponds within errors to the mean charge for a single photo electron.

The 8 bit amplitude resolution in the test setup was somewhat limiting the resolution of the single photo electrons due to the LONS. With the higher 
resolution of 10 bit with the full MUX-FADC system even a continuous calibration 
of the read-out chains using the single photo electrons shall be possible.

\subsubsection{MC Simulations}

The response of the MAGIC telescope to gamma ray showers and to background was 
simulated in detail \cite{Magic_MC}. Both the currently used 300 MSamples/s 
readout chain and the ultra-fast digitization were simulated.

\begin{figure}[h!]
\begin{center}
\includegraphics[totalheight=7cm]{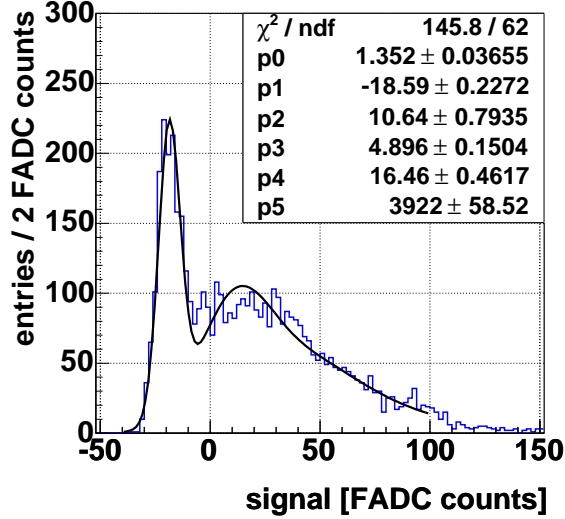}
\end{center}
\caption[Reconstructed single photo electron spectrum of a MC pedestal run.]
{\small \it Reconstructed single photo electron spectrum of a simulated pedestal 
run. The leftmost peak is the pedestal.} \label{fig:single_ph_el_MC_fit}
\end{figure}

Figure \ref{fig:single_ph_el_MC_fit} shows the reconstructed single photo 
electron spectrum of a simulated pedestal run. The highest integral of 8 FADC 
slices (4~ns) was searched for in a fixed digitization window of 20~slices 
(10~ns). The leftmost peak corresponds to electronics noise only. The right part 
of the distribution corresponds to the response of the PMT to one or more photo 
electrons.

\begin{figure}[h!]
\begin{center}
\includegraphics[totalheight=6cm]{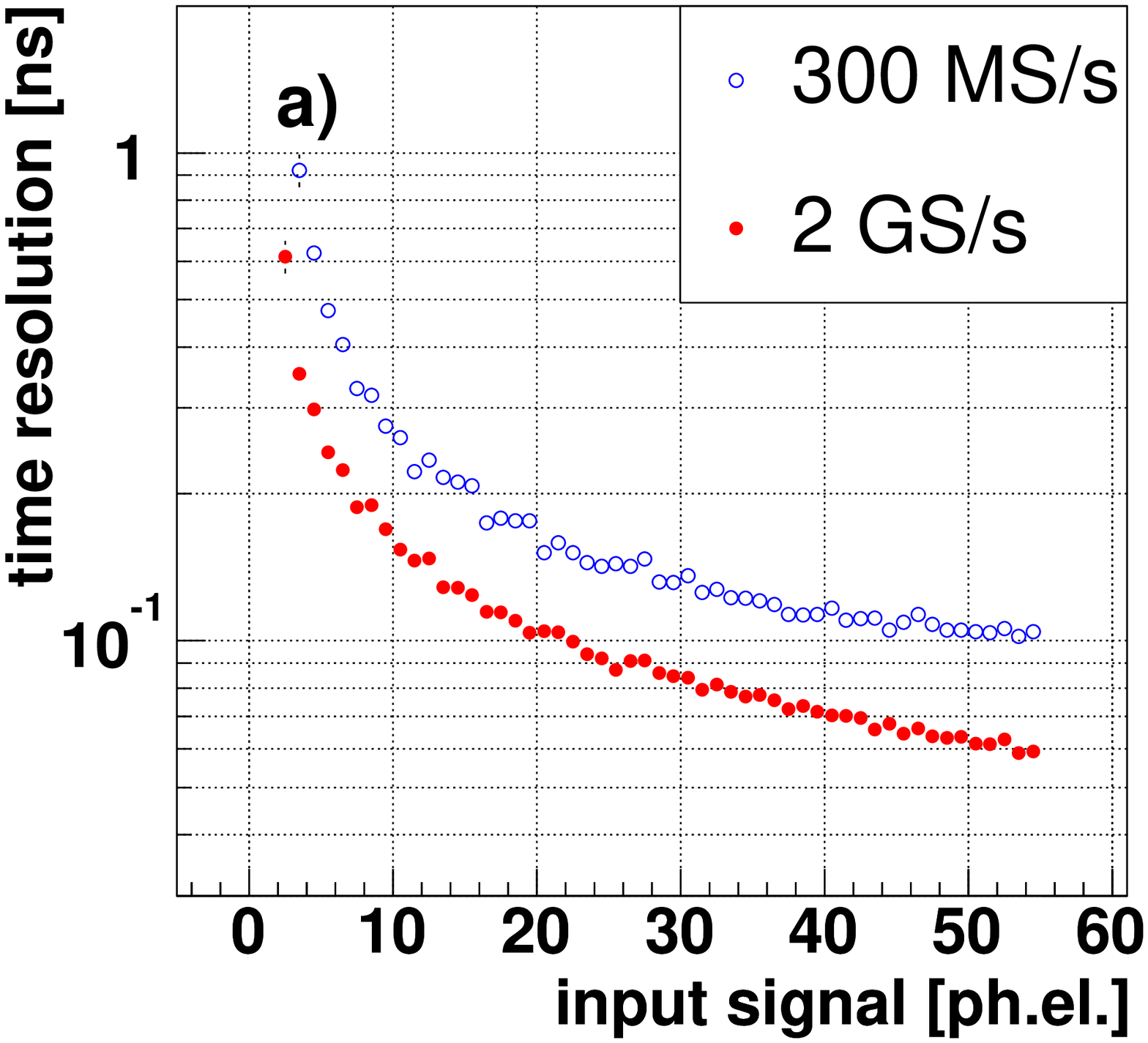}
\includegraphics[totalheight=6cm]{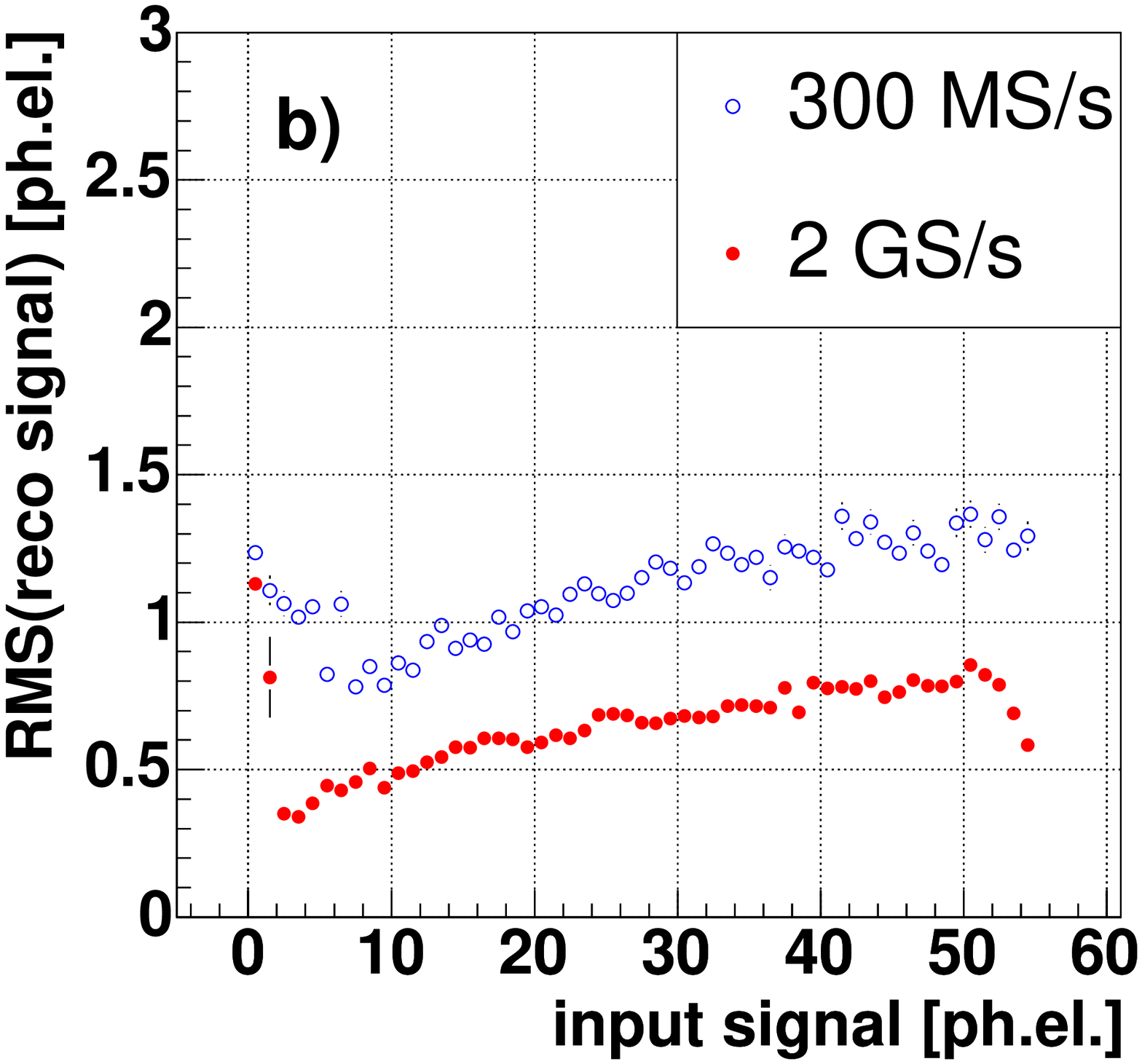}
\end{center}
\caption[Comparison of the pulse arrival time resolution.]{\small \it MC 
simulations: a) Comparison of the pulse arrival time resolution as a function of 
the input signal size between the current MAGIC 300 MSamples/s FADCs and the 2 
GSamples/s FADCs. The time resolution improves by nearly a factor of 2 with 
the new system. b) Comparison of the signal resolution as a function of the 
input signal with the current 300 MSamples/s MAGIC FADCs and the 2 GSamples/s 
FADCs system. The signal resolution improves by a factor of about two.} 
\label{fig:time_resolution_old_new}
\end{figure}

In figures \ref{fig:time_resolution_old_new} the signal and arrival time 
resolutions of the current and the MUX-FADC system are 
compared using MC simulations. For both MC simulations the same LONS conditions are assumed as well as the same electronics noise level. In the simulation no intrinsic transit time spread of the PMTs of about 250 ps 
per photo electron was taken into account. The input light pulse has a FWHM of 1~ns as expected for gamma-ray induced showers.

Contrary to the simple signal and arrival time extraction algorithms used above, a dedicated numerical fit to the FADC samples using a known pulse shape \cite{OF94,OF_Magic} has been applied to illustrate the theoretically achievable resolution. Figure \ref{fig:time_resolution_old_new}a shows the resolution of the reconstructed 
pulse arrival time as a function of the input signal. The MUX-FADC system 
improves the timing resolution by nearly a factor of 2. Figure \ref{fig:time_resolution_old_new}b shows the 
resolution of the reconstructed charge as a function of the input charge. With 
the MUX-FADC system the charge resolution improves by a factor of two.

\section{Discussion}\label{sec:discussion}

The ultra-fast fiber-optic multiplexed FADC prototype read-out system was 
successfully tested during normal observations of the MAGIC telescope in La 
Palma. The fiber-optic splitters and delays are commercially available and 
comply with the required specifications for the use in the ultra-fast MUX-FADC 
read-out system. The 10 bit 2 GSamples/s FADCs from Acqiris have been developed 
for MAGIC, and are available now as a commercial product. Thus the ultra-fast 
FADC 
read-out has grown to a mature technology which is ready for the use as a 
standard read-out system for the MAGIC telescope and other high-speed data 
acquisition applications.

The multiplexing of 16 channels into one ultra-fast FADC allows one to greatly 
reduce the price of an ultra-fast read-out system. The MUX-FADC read-out reduces 
the costs by about 85\% compared to using one ultra-fast FADC per 
read-out channel. Also the power consumption of the read-out system is greatly 
reduced.

The ultra-fast MUX-FADC system allows to skip the artificial pulse stretching and thus to use a  shorter integration window for the Cherenkov pulses. The reduction of the pulse integration window from 13.33~ns (4 samples with 3.33 ns per sample) for the current MAGIC FADC system to 5~ns (10 samples with 0.5 ns per sample) for the MUX-FADC system corresponds to a reduction of the integrated LONS charge by a factor of about 2.7. Consequently, the RMS noise of the LONS is reduced by about 40\%.


The recorded images of the air showers are usually, at least for energies above 100 GeV, characterized by so-called Hillas parameters \cite{Hillas_parameters}. In order to prepare raw shower images for the Hillas parameter calculation it is necessary to apply so-called tail cuts to reject pixels with a low signal to noise ratio. All pixels with signals below 3 times their noise RMS (mainly due to LONS) are rejected (dynamic image tail cut cleaning).

A reduction in the noise RMS translates into lower image cleaning levels. Thus a larger part of the shower image, or in other words a shower image of a higher signal to noise ratio, can be used to calculate Hillas parameters. This is especially important for low energy events where the signals of only a few pixels are above the image cleaning levels. The image quality of the air showers will improve with the ultra-fast read-out system. This will allow the reduction of the analysis energy threshold of the MAGIC telescope. 

The ultra-fast FADC system also provides an improved resolution of the 
timing structure of the shower images. As indicated by MC simulations \cite{muon_rejection} gamma 
showers, cosmic ray showers and the so called single muon events have different 
timing structures. Thus the ultra-fast FADC read-out can enhance the separation 
power of gamma showers from backgrounds.

After the successful prototype test of the ultra-fast MUX-FADC read-out system 
it is ready to be installed as a future read-out of the MAGIC telescope.


\appendix

\section{Acknowledgements}

The authors thank R. Maier and T. Dettlaf from the electronics workshop of MPI 
for the layout, design and production of the MUX-FADC electronics. We also 
acknowledge the very good collaboration with the companies Acqiris and 
Sachsenkabel.





\end{document}